# A systematic review of recent air source heat pump (ASHP) systems assisted by solar thermal, photovoltaic and photovoltaic/thermal sources

Xinru Wang [a], Liang Xia [a, *], Chris Bales [b], Xingxing Zhang [b, **], Benedetta Copertaro [b], Song Pan [c], Jinshun Wu [d]

[a] Research Centre for Fluids and Thermal Engineering, University of Nottingham Ningbo China, China

[b] Department of Energy and Built Environment, Dalarna University, Falun, Sweden

[c] Engineering Research Centre of Digital Community, Beijing University of Technology, Beijing, China

[d] College of Architecture & Civil Engineering, North China Institute of Science & Technology, Hebei, China

**Abstract**:

The air source heat pump (ASHP) systems assisted by solar energy have drawn great attentions, owing to their great feasibility in buildings for space heating/cooling and hot water purposes. However, there are a variety of configurations, parameters and performance criteria of solar assisted ASHP systems, leading to a major inconsistency that increase the degree of complexity to compare and implement different systems. A comparative literature review is lacking, with the aim to evaluate the performance of various ASHP systems from three main solar sources, such as solar thermal (ST), photovoltaic (PV) and hybrid photovoltaic/thermal (PV/T). This paper thus conducts a systematic review of the prevailing solar assisted ASHP systems, including their boundary conditions, system configurations, performance indicators, research methodologies and system performance. The comparison result indicates that PV-ASHP system has the best techno-economic performance, which performs best in average with coefficient of performance (COP) of around 3.75, but with moderate cost and payback time. While ST-ASHP and PV/T-ASHP systems have lower performance with mean COP of 2.90 and 3.03, respectively. Moreover, PV/T-ASHP system has the highest cost and longest payback time, while ST-ASHP has the lowest ones. Future research are discussed from aspects of methodologies, system optimization and standard evaluation.

## Content

**Abbreviations**

| | |
|---|---|
| ASHP | Air source heat pump |
| (BI)PV | (Building integrated) Solar photovoltaic |
| (BI)PV/T | (Building integrated) solar Photovoltaic/thermal |
| COP | Coefficient of performance |
| DA-SA | Direct expansion solar assisted |
| DHW | Domestic hot water |
| DMHP | Dual source multi-functional heat pump |
| DX-SA-ASHP | Direct solar assisted air source heat pump |
| EER | Energy efficiency ratio |
| EEV | Electronic expansion valve |
| EIF | Emission intensity factor |
| EPBT | Energy payback time |
| EROI | Energy returned in invested |
| FPC | Flat plate collectors |
| GHG | Greenhouse gases |
| HE | Heat exchanger |
| HP | Heat pump |
| HVAC | Heat, ventilation & air conditioning |
| IDX-SA-(ASHP) | Indirect solar assisted (air source heat pump) |
| IX-SAMHP | Indirect expansion solar-assisted multi-functional heat pump |
| LCA | Life cycle assessment |
| LCC | Life cycle cost |
| PCM | Phase change material |
| PV | Photovoltaic |
| PV+AC | Photovoltaic + air conditioning unit |
| PV-ASHP | Photovoltaic assisted air source heat pump |
| PV/T | Photovoltaic/solar thermal |
| PV/T-ASHP | Photovoltaic/solar thermal assisted air source heat pump |
| PV-LHP | Photovoltaic loop heat pipe |
| RFHW | Floor heating without water |
| SAHP | Solar assisted heat pump |
| SF | Solar fraction |
| SIASHP | Solar integrated sir source heat pump |
| SPF | Seasonal performance factor |
| SRCD | Standard reverse-cycle defrosting |
| ST | Solar thermal |
| ST-ASHP | Solar thermal assisted air source heat pump |
| TCC | Tolerable capital cost |
| TES | Thermal energy storage |
| TRESE | Triple-sleeve energy storage exchanger |
| TRTHE | Triple tube heat exchanger |
| TS VC | Two-stage variable capacity |
| TTCC | Total tolerable capital cost |
| WH | Water heater |
| WSHP | Water source heat pump |

# 1 Introduction

## 1.1 Background

Energy saving has become one of the most important subjects as energy shortage is getting worse and the demand for energy is rising rapidly worldwide in recent decades [1]. Compared with rational boiler central heating or electrical heating, heat pump is a more efficient and environment friendly system to supply suitable indoor climate [2]. As a result, it earns a dominant position in the air condition market. Especially in recent years, coal fired boiler heating is identified as the major cause of catastrophic air pollution in many cities of China, such a Beijing [3]. Among all types of heat pump systems, Air Source Heat Pump (ASHP) is a state of art heating system with many advantages, such as low energy consumption, relatively stable performance, huge energy-saving potential [4,5], abundant social benefits [6,7], and environmental-friendly characteristics [8,9]. ASHP systems have already been widely applied in different types of buildings [10,11]. However, the performance of different ASHP systems are affected by complex outdoor conditions, especially in cold regions [12,13]. To be more specific, performance of ASHP systems drop significantly under the low temperature environment and even get frosting under the worst scenario [14,15]. There are many technical solutions to improve the ASHP performance, such as secondary suction [13], adding insulation to evaporator [14] and using auxiliary source [15, 16] and so on. These methods have certain limitations, as they can only help to improve the performance in a limited extent especially under severe climate conditions. Moreover, some methods need consume much more energy to guarantee the high performance, such as using auxiliary heat. To solve these problems further step, solar assisted ASHP system is proposed as one solution, which is capable of supplying solar energy to the ASHP system by using direct solar irradiation and latent heat from the solar and air. While air and solar energy are both renewable and clean energy meeting the energy consumption trend that all over the world encourage to use the sustainable and renewable energy. Solar assisted ASHP are hybrid systems, which can be coupled with solar thermal (ST), photovoltaic (PV) array, or both photovoltaic/solar thermal (PV/T), abbreviated as ST-ASHP, PV-ASHP and PV/T-ASHP in this paper [21]. The major benefits of solar assisted ASHP systems include that: (1) these systems can achieve higher techno-economic and environmental performance by improving the operation of ASHP systems [17,18,19, 20,21]; (2) they can also meet higher ratio of the energy demand [22,23,24,25,26,27]; and (3) they can produce different types of energies, which is not only heating/hot water or cooling, but also possible electricity from PV panels [28,29,30,31,32]. As a result, this paper will concentrate the R&D work of three most prevailing solar assisted ASHP systems, including solar thermal assisted ASHP (ST-ASHP), photovoltaic combined ASHP (PV-ASHP) and photovoltaic/thermal (PV/T-ASHP).

### 1.2 Scope and objectives

There has been a growing interest for the solar assisted ASHP systems with a variety of system configurations for various climate conditions. A large number of research studies have been conducted in the literature on fundamentals of system design/optimization, performance modelling, experimental testing and pilot applications. This, in turn, increases the degree of complexity for users to compare and implement the systems.

Most of the solar assisted ASHP systems have complex design and working principles dedicating to certain applications in specific climate or buildings. Their corresponding performance therefore are diverse, depending on boundary conditions and characteristic parameters. It is not clear enough which system to use in a certain boundary condition and what are the according performance. A knowledge gap thus needs to be filled in by a review work.

This study thus aims to tackle with such challenge by presenting a systematic review of the recent R&D works on above-mentioned three solar assisted ASHP systems for comparison of their performance. The associated research methods, application and main results of these systems are subsequently analysed.

The specific objectives of this review paper are designed to: (1) provide a systematic review on solar assisted ASHP system with their working principles; (2) evaluate the performance of the three-mentioned solar assisted ASHP systems; (3) depict the methodologies and results/characteristic parameters in various boundary conditions and compare the advantages and disadvantages of solar energy assisted ASHP systems, and (4) illustrate the feasible applications in practice.

### 1.3 Method and structure

This paper is based on existing studies found in scientific publications from peer reviewed journals, conferences searched in Google scholar and Web of Science. Some of the figures presented in this paper are converted from or calculated based on data in the reviewed publications to make it easier for comparison. To limit the review, only air source heat pump combined ST, PV and PV/T are considered. This means other energy source heat pump systems are not included, e.g. water source heat pump, ground source heat pump. The entire article focused on the system level and organized as follows: section 2 introduces the system boundaries and assessment indicators of solar energy assisted ASHP systems. Section 3 and 4 reviews the existing researches in recently near ten years papers. An overall performance comparison of the systems is give in section 5. In section 6, promising research directions for enlarging popular market application are discussed. Finally, salient conclusions are drawn in section 7.

## 2 System boundaries and performance indicators

### 2.1 System boundaries

System boundaries are important to understand the system and evaluate their performance in an equivalent condition. While in many studies, researchers do not clearly define the system boundaries of solar assisted ASHP systems, which makes it difficult to distinguish the different systems. Within IEA SHC Task 44/HPP Annex 38 (T44A38), the various system boundaries of solar and heat pump systems were introduced [33], shown in Fig.1. Heat sources are identified in green, while purchased energy electricity is in grey. The red boxes on the right represent the energy supplied to users.

For conventional ASHP system, the heat pump gains heat from air and electricity from grid. In theoretical working principle, conventional ASHP system can supply heating, cooling and hot water for users, but with the climate limitations, ASHP systems generally need auxiliary heaters to meet the user demand in severe conditions. Solar assisted ASHP system can use both air and solar energy as the energy sources. The system configuration is explained in section 2.2 and the definitions of system boundaries shown in Fig. 1. There are also other sources heat pump system, like ground, water and so on, in this paper, we only show the air source heat pump system.

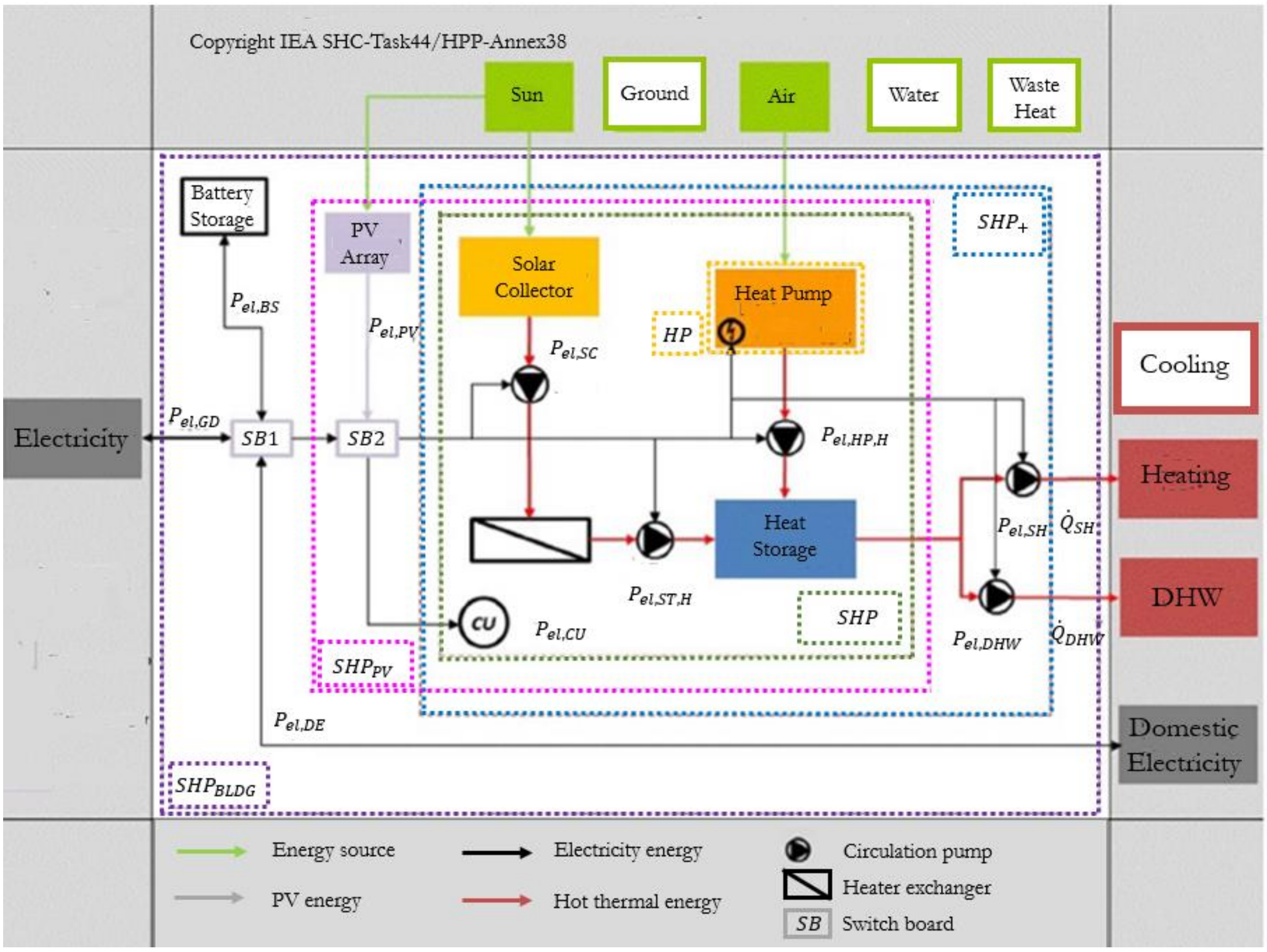

Fig. 1 Different system boundaries (figure based on the ongoing Task 53 [34] work).

According to the Task 53 [34], the definitions of solar assisted air source heat pump system boundaries (Figure 1) mainly contain five types. The origin air source heat pump system is shown in orange dashed line (HP), which contains heat pump only. Adding solar collector system, which is concluded in green dashed line, the system is called solar assisted air source heat pump system (SHP), while above two system types are both without circulation water pump. System boundary SHP+ does include these pumps (blue dashed line). Solar assisted air source heat pump system with PV arrays ($SHP_{PV}$) is shown in red dashed line. It includes the PV array but without battery storage. With battery storage system, the system boundary is called solar assisted air source heat pump system ($SHP_{BLDG}$) for building heating, DHW and electricity (purple dashed line) [35]. Corresponding to Task 53, in this paper, SHP+ is recorded ST-ASHP system, $SHP_{PV}$ boundary is the PV-ASHP system and $SHP_{BLDG}$ is the PV/T-ASHP system.

## 2.2 System configuration

There are many different ways of categorization of solar assisted ASHP system. In Fig.1, the system boundaries are identified on a component level, e.g. heat pump, or on a system level. Among them, the solar heat pump ($SHP_+$) system includes the circulation pumps, while SHP and HP does not, which is used for comparing the assessment of the system environmental impact in operation [33]. Besides that, they can also be divided based on the type of heat demand to be served, e.g. heating, DHW [36]. In addition, the interaction way of the heat pump and the solar assisted system is another common categorization. This paper categorizes the solar assisted ASHP systems according to the evaporator component, including ST-ASHP, PV-ASHP, and PV/T-ASHP.

### 2.2.1 ST-ASHP system

For ST-ASHP system, the solar thermal collector convert solar irradiation into thermal energy that could be used directly by users or supply to ASHP. ASHP produce heat/cold for users. ST-ASHP system could supply different types of energy such as heating, cooling, hot water even the steam.

Based on the generation method, the ST-ASHP system can be categorized into direct expansion solar assisted (DX-SA) and indirect expansion solar assisted system (IDX-SA). In DX-ST-ASHP, the heat pump and solar thermal system work together as one combined system and solar thermal collector acts as the system evaporator. Indirect expansion solar thermal and air source heat pump (IDX-ST-ASHP) systems can generally be classified in parallel, serial, regenerative (Fig.2) and complex system concepts by the interaction between solar thermal and heat pump [33, 37]. The serial type is common in solar and ground source heat pump combination system, while for solar thermal assisted air source heat pumps system, a statistical analysis on market-available ST-ASHP systems can be found in Ruschenburg et al. [38,39]. Their main result indicated that the parallel systems are the market-dominating system concepts (61% of the surveyed systems) [39]. In parallel configuration (Fig.2 (a)), solar collector and air source heat pump independently supply heating and/or DHW, or cooling to users.

ST-ASHP system commonly use two types of media: liquid and air. Based on media temperature, ST-ASHP is divided into two types. Low and medium temperature is less than 100℃, 250-2500℃ is high temperature types. While the common temperature used in real application of ST-ASHP system usually is the low and medium temperature types.

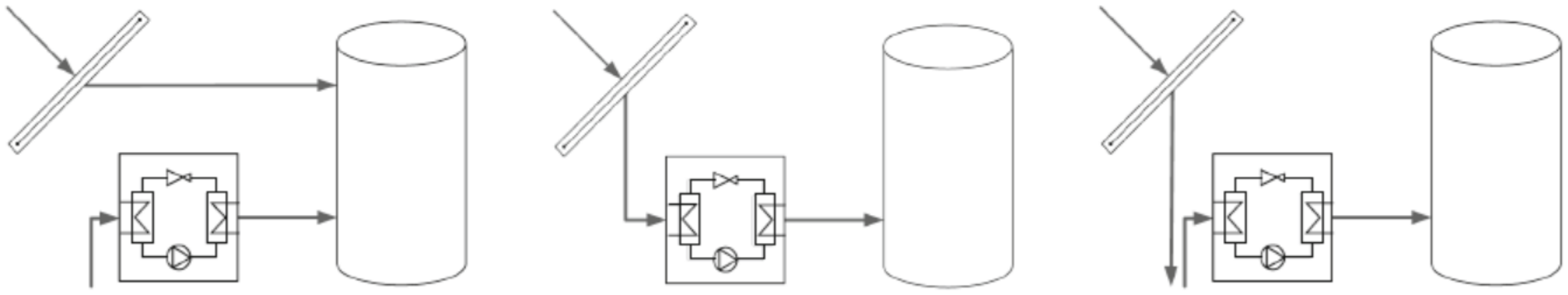

Fig. 2 Classification of solar thermal and heat pump systems in parallel, serial and regenerative system concepts: (a) parallel; (b) serial; (c) regenerative [37]

### 2.2.2 PV-ASHP system

Solar photovoltaic (PV) [40] can direct conversion of sunlight into electricity without any heat engine to interfere. The ASHP partially covered by the electricity produced by PV in the buildings is called PV-ASHP system. Nowadays, there is also another popular type of the PV which integrated with the building such as the roofing, window and so on, called building-integrated photovoltaic (BIPV) systems. PV-ASHP system could supply heating, cooling and electricity to buildings. The principle of PV/T-ASHP system is that the electricity produced by the PV can be supplied to the ASHP which supply the heating or cooling to the building, when the PV cannot meet the demand, the ASHP can gain electricity from the grid. Meanwhile, the overcome electricity produced by PV can also be added to the grid in the electricity producing time, e.g. midday in the sunny days when the load of building is also small.

PV-ASHP system could be classified direct expansion and indirect expansion types. Moreover, it can also be classified by the array types [27]. While the common classification is based on the lighting absorbing materials. The most common material is silicon which include silicon, amorphous silicon and crystalline silicon and there are other materials such as cadmium telluride and cadmium sulphide, organic and polymer cells and hybrid photovoltaic cell and some other cells.

### 2.2.3 PV/T-ASHP system

Normally, PV module only converts 4-17% of incoming solar radiant into electricity, which varies along with the material and working conditions, and as a result, much more solar energy is converted into heat. Over high ambient temperature could reduce the efficiency of the PV and cause damage to the structure to the modules [54]. PV/T can produce electricity as well as heat, which can cool the surface temperature of cells. Like the BIPV system, the PV/T integrated with building called BIPV/T system, which can save the space than two separated arrays of PV and solar thermal collectors. Combined with ASHP, PV/T-ASHP system can supply heating and domestic hot water (DHW), even in the future cooling for buildings.

While in PV/T-ASHP system, as PV/T collectors can not only provide heat, but also produce electricity, the common type is in serial connection, as the PV/T collectors are used to regenerate the heat exchanger [41]. PV/T systems can be divided into five types: air cooled system, water cooled system, both water and air cooled, using PCM PV/T system and using heat pipes in PV/T system [42]. And more recently, using PCM in PV/T system is concentrated [43]. And based on the integrated fabric of the building, the BIPV/T system can be divided into wall integrated, window integrated, roof integrated and façade integrated systems. For the whole PV/T-ASHP system, according to the application, there are heating, DHW intents types.

## 2.3 System performance indicators

With a variety of goals and system boundaries, there are many different performance indicators to evaluate the solar assisted ASHP systems. In this section, the performance of solar assisted ASHP system is regarded from a boarder concept including energy, economy and environment aspects, within certain boundaries under given operating conditions for a defined time.

### 2.3.1 Energetic performance indicators

Coefficient of performance (COP) and energy efficiency ratio (EER) are usually used as the energetic performance indicators for heat pump unit under defined operating conditions, as the EER is used for cooling application in European [33]. While seasonal performance factor (SPF) is used to express the overall system efficiency including all auxiliary components, e.g. storages. There are also other figures, e.g. collector thermal efficiency, and the component performance figures such as thermal output of collector, in this paper, they are not listed, as they are not common in the reviewed solar assisted ASHP system papers.

COP or EER [44,45] is evaluated by measuring the temperature of the inlet and outlet of the heat transfer media and the electricity consumption, as written in Eq.(1).

$$COP_{sys} = \frac{Q_t}{P_{el}} \quad (1)$$

Where, $COP_{sys}$ is the overall performance of system, $Q_t$ is the heat gained in system, $P_{el}$ is the power consumption.

System energy efficiency ratio ($EER_{SYS}$) can be calculated from the energy efficiency ratio of the unit ($EER_{UNIT}$) and the Solar Contribution (SC) by Eq. (2).

$$EER_{SYS} = EER_{UNIT} \frac{100 - SC(\%)}{100} \quad (2)$$

Seasonal performance factor (SPF) is the ratio of overall useful supplied energy to users to the overall electricity used during operation, which is defined in Eq. (3). Where. $Q_{SH}$ is the heat amount for space heating, $Q_{DHW}$ is the heat amount for domestic hot water.

$$\text{SPF} = \frac{Q_{SH} + Q_{DHW}}{P_{el}} \quad (3)$$

EER is used for cooling application, while for heating and hot water, COP and SPF are commonly used. So the definition of COP and SPF not including EER in Eq. (1) and (3) for electricity for different system boundaries is concluded in Table 1. However, there are very limited information that can be found on the EER, COP or SPF at building level in existing literatures, which thus need more attention in the future.

Table 1 Definition of SPF/COP and the total power for different system boundaries in Fig.1 [35]

| System boundary | Definition of COP/SPF | Total Electrical power |
|---|---|---|
| HP | $COP_{HP}/SPF_{HP}$ | $P_{el} = P_{el,HP} + P_{el,AS}$ |
| SHP | $COP_{SHP}/SPF_{SHP}$ | $P_{el} = P_{el,HP} + P_{el,AS} + P_{el,SC} + P_{el,ST,H} + P_{el,HP,H} + P_{el,CU}$ |
| $SHP_{PV}$ | $COP_{SHP,PV}/SPF_{SHP,PV}$ | $P_{el} = P_{el,HP} + P_{el,AS} + P_{el,SC} + P_{el,ST,H} + P_{el,HP,H} + P_{el,CU} - (P_{el,SHP} \cap P_{el,PV})$ |
| $SHP_{+}$ | $COP_{SHP+}/SPF_{SHP+}$ | $P_{el} = P_{el,HP} + P_{el,AS} + P_{el,SC} + P_{el,ST,H} + P_{el,HP,H} + P_{el,CU} + P_{el,SH} + P_{el,DHW}$ |

Where, electricity use $P_{el,HP}$ and $P_{el,\mathrm{AS}}$ are the total electrical energy use of the heat pump compressor and fans (system boundary HP). $P_{el,SC}$ is the total electrical energy use of solar circuit, $P_{el,EH}$ is the total electrical energy use of auxiliary electrical heater (if included, not in this case). $P_{el,CU}$ is the total electrical energy use of control units, while the others are all circulation pump consumption. $P_{el,\mathrm{ST,H}}$ and $P_{el,\mathrm{ST,C}}$ are the total electrical energy use of hot and cold storage circulation pumps. $P_{el,\mathrm{SH}}$ and $P_{el,DHW}$ are electric space heat and DHW circulation pump consumption. $P_{el,SHP} \cap P_{el,PV}$ is the intersection electric power of heat pump consumption and PV generation.

### 2.3.2 Economic performance indicators

In terms of economic aspect, the capital investment is always considered. Christos et al [46] compared the net present value of all investment including the capital cost and the electricity consumption of solar assisted ASHP with WSHP and fan coil system. Evangelos et al [47, 48] also considered the total cost to examine the system. Besides investment, the payback time, the energy saving also compared by [47].

Life cycle cost (LCC) in Eq.(4) are used to compare different systems. Stefano Poppi et al [49] developed an economic model based on a comparative cost-analysis between the system variations and the reference system for simulation.

$$\mathrm{LCC} = C_{system} + C_{installation} + C_{maintenance} + C_{energy\ usage} \tag{4}$$

Accurate estimation of solar assisted ASHP system's initial investment is difficult because the installed costs can vary significantly depending on the scope of the plant equipment, geographical area, competitive market conditions, special site requirements, and prevailing labour rates. Thus, an alternative approach to conventional economic feasibility analysis is adopted, which involves the calculation of the "tolerable capital cost" (TCC) of the upgrades [50]. TCC is the capital cost for an energy saving upgrade that will be recovered based on the annual savings, the number of years allowed for payback, and the estimated annual interest and fuel cost escalation rates.

Other papers concentration on the payback time, and the energy payback time (EPBT) [51,52] and energy returned on energy invested (EROI)[45] for the solar assisted ASHP life cycle assessment (LCA). LCA is a methodology to assess the environmental impact of a product or service over its whole life cycle, which may include processes from the extraction of raw materials to recycling or disposal [53].

$$\mathrm{EPBT(year)} = \frac{Embedded\ (primary) energy\ (MJ/m^2)}{Annual\ (primary) energy\ generated\ by\ the\ system\ (MJ/(m^2 \cdot year))} \quad (5)\ [48][49]$$

$$\mathrm{EROI} = \frac{\text{lifetime energy output}}{\text{Embedded energy}} \quad (6)[49]$$

However, compared with energetic performance indicators, the economic performance indicators are less popular. Most of the papers that mentioned the economic performance only simplified compared the investment or showed the payback time. Moreover, different researches referred different contents on economic aspect, which is difficult to make a comparison among different systems.

### 2.3.3 Environmental performance indicators

For environment impact, some papers [54,55] analysed the greenhouse gases (GHG) especially $CO_2$ emission reduction. The GHG emissions are evaluated and reported as "equivalent $CO_2$" ($CO_{2e}$) emitted per unit input energy, which is determined based on the fuel type and efficiency of the energy conversion devices. The GHG emission associated with electricity use is determined based on the amount of fossil fuel consumption to produce and deliver electricity to a building, which is updated in each time step using the actual fossil fuel consumption [56]. While $CO_2$ emission by wood combustion returns to the atmosphere where the $CO_2$ that was recently removed by photosynthesis as the tree grew [57].For comparison among different systems, the GHG emission intensity factor (EIF) is defined. The GHG EIF is defined as the level of $CO_2$e emission for generation and delivery of 1 kWh electricity to the end-user. While, usage need to take into account the different electricity generation ways during peak and base periods. Thus, different average and marginal GHG EIFs are developed to address electricity generation within the base and peak periods [57]. Moreover, besides $CO_2$, the greenhouse gas emission of PV-ASHP like $SO_2$ and $NO_x$ emission was also calculated in some papers [**Error! Bookmark not defined.**85].

Although there are some researches noticed the environmental performance of solar assisted system, the number is few, and most of them only mentioned these results in few sentences. There are a few reason, the first is that there is no standard calculation method for all researches to make a comparison. Second, the measurement for different systems are also different, as different authors focus on different aspects on their own system.

## 3 Simulation researches and the related results

Table 2 summarizes the main research findings of simulation in solar assisted ASHP, including the tools applied, the built model types, the corresponding models' operation conditions, the input and output parameters and their main results [58].

The simulation models for solar assisted ASHP system include TRNSYS, CFD, self-developed Matlab and mathematic model, in which TRNSYS is dominant. The second most used models are built on various mathematical equations.

In the TRNSYS models, type 56 is the most model for multi-zone buildings and type 832 is commonly used for collector. Heat exchanger is built by different types including type 805, type 91 and type 557a. Heat pump models are type 401 for compressor heat pump, type 655-3, type 665, type 887, and type 841 for $CO_2$ heat pump. For store devices, type 340, type 534 and type 4 are used. Besides, type 15 is applied for weather data reading and processing, type 6 is used for electrical heater.

The operation conditions can be classified by outdoor environment including air temperature, solar irradiation and operation mode including heating mode, cooling mode and DHW, while for ASHP with phase change material (PCM-ASHP) system, operation modes include charging/discharging mode.

For the input parameters, the surrounding conditions [59], including the location which means the solar and ambient air temperature, are the most important parameters. Moreover, some researchers highlight the components' design and operation parameters. To evaluate the solar assisted ASHP system, COP, SPF, payback time and GHG are often used to assess the system performance.

In the following subsections, we will discuss the detailed simulation R&D status of each type of solar assisted ASHP systems, mainly by choosing several representative cases from this table.

Table 2 Summary of simulation methods and the related results of solar assisted ASHP systems

| Ref. | System | Tool | Model | Location | Operation Conditions | Input data | Building Types | Evaluation parameters | Results |
|---|---|---|---|---|---|---|---|---|---|
| [60] [49] [61] | parallel ST-ASHP | TRNSYS 17 | Type 832 QDT multimode model for collector; Type 340 for store; Type 805- heat exchanger; Type 887-heat pump | Zurich and Carcassonne, Sweden | SFH100 in Carcassonne | location and house model; interest rate, inflation rate and price of electricity; | laboratory | SPF; Total electricity use; Annual DHW discharge energy; collector area; tank volume; UA-value of DHW heat exchanger; Payback time; | ▪SPF= 2.84; ▪EPBT= 10 years ▪ system electricity change between 305 and 552 kW h/y when collector area from 5 to 15 $m^2$; ▪main influence for system including annual DHW discharge energy, collector area, UA-value of DHW heat exchanger and heat pump size; |
| [62] | Different SAHP (review) | TRNSYS | Type 56-ASHP; Type 655-3-default heat pump | — | — | location and equipment parameters; | — | energetic comparison; indoor temperature distribution; financial evaluation; | ▪ASHP system cost lowest investment and highest electricity among solar assisted WSHP and fan coil heating system; |
| [63**Error! Bookmark not defined.**] | parallel ST-ASHP/ST-GSHP | TRNSYS | Type 56-building; Type 832-collector; Type 401-compressor HP; Type 557a-borehole heat exchanger; Type 340-multiport storage tank | Helsinki, Strasbourg, Athens | solar heating/ $CO_2$ heat pump heating mode | Location; | building type SFH45 (TRNSYS) | SPF; | ▪The design of SHP systems depending on the boundary conditions of the specific application; ▪COP of ASHP= 3.66 |
| [64] | solar combined $CO_2$ heat pump | TRNSYS | Type 71-solar collector; Type 841-$CO_2$ heat pump; Type 4- storage tank; Type 534- operation tank; | Shanghai, China | — | Taiyuan, Shanghai | residential building | volume of storage tank and operation tank; | ▪the optimization values: storage tank volume=2.21 $m^3$ and operation tank volume= 0.3 $m^3$ ▪the optimized system can save 14.2% electricity and solar fraction can reach 71.1%. |
| [65] | PV-LHP/SAHP | mathematical model | — | Qinhuangdao, China | solar/air source heat pump | solar irradiation and ambient air temperature | residential buildings | operation performance of under typical working conditions; economic feasibility analyses; | ▪solar energy utilization efficiency decreased with solar irradiation growth and ambient temperature decrease; ▪ monthly average COP of heat pump modes= 3.10; ▪The annual solar heating ratio of the system is up to 57.8%. ▪ Life cycle cost of the PV-LHP/SAHP could be reduced by 29.6% than ASHP. |
| [66] | ST-ASHP | TRNSYS 17 | Type 1-quadratic efficiency collector; Type 3- pump; Type 2b-controller; Type 15-weather data reading and processing | Taipei and Kaohsiung | — | setting parameters for components | lab-scale system | SPF and solar fraction; Payback time; | ▪The overall efficiency of a DHW system improved, while the cost is also increased; ▪ EPBT= 5 years; ▪ SPF= 4.56 / 4.93 in Taibei and Kaohsiung |
| [67] | DX-SHPWH | mathematical model | — | Xi'an, China | the solar radiation ranges; ambient air temperature superheating degrees of the two evaporators; | component parameters | -- | heating capacity; COP; | ▪Solar irradiation and ambient air temperature have a great effect on the system performance; ▪The COP on average increases by 14.1%; ▪Given solar radiation of 500W/m2, the COP=4.98 and Qc = 605 W |
| [68] | ST-ASHP | TRNSYS | one-node model | Geneva (Switzerland). | (i) direct solar heat production; (ii) storage discharge, (iii) activation of the HP, with surplus production (iv) direct electric heating | hourly weather data; hourly load demand; | multifamily buildings | SPF; | ▪SPF of system achieved maximize 4.4 and when temperature distributed, the SPF is 3.1-4.1; ▪SPF of system achieves 5 is potential ▪investment and roof area reality condition are not considered; |
| [69] | PV-ASHP; PVT-WSHP; | TRNSYS | Type 56-building; Type 655-3-heat pump; | Greece | heating mode | undefined parameter; the collecting area and | commercial building | monthly load; COP; energy consumption; | ▪lower energy consumption than ASHP and fan coil heating system; ▪Indoor temperature is better in solar driven heat pump system; |
| | | | | | | storage tank volume | | | ▪The COP is near 4 and for the conventional air source systems close to 2.5. |
| [70] | ASHP vs. ASHP BIPV/T | TRNSYS | Type 568-BIPV/T system;<br>Type 56-multi-zone building;<br>Type 687-a window model; | Anchorage, southern Alaska, USA | the ambient air supplied to HP<br>the warm air coming out of BIPV/T was supplied into HP | irradiation | residential building | COP; | ▪COP improved for average ambient temperature above -3° C;<br>▪COP did not be improved, when average ambient temperatures were above 10 ° C or below -10 ° C;<br>▪ The maximum COP of BIPV-ASHP= 5.31; |
| [71] | BIPV/T-ASHP | TRNSYS | — | Toronto, Canada | cooling/heating/DHW/floor heating | air flow rate | test hut | COP; | ▪An integrated BIPV/T+ASHP thermal energy system increases the overall performance;<br>▪the seasonal COP could be increased from 2.74 to 3.45.<br>▪The heat pump electricity consumption is reduced by 20% for winter. |
| [72] | BIPV/T-ASHP | TRNSYS 17 | Type 567-BIPV/T;<br>Type 56-building model;<br>Type 665-ASHP; | Ontario, Canada | heating mode | outdoor temperature | Archetype Sustainable House | COP;<br>saving in energy and cost;<br>GHG emission reduction; | ▪ The heat pump electricity consumption is reduced by 20% for winter<br>▪GHG emission reduced and energy consumption got saving;<br>▪COP increased from 2.74 to 3.45; |
| [73]<br>[74] | PV/T-ASHP | CFD | Gambit;<br>Fluent | Shenyang, China | steady state RNG k-ε turbulence model | inlet and outlet velocity | residential building | efficiency of BIPV/T;<br>COP; | ▪ COP of heat pump unit reached 4.6;<br>▪Thermal efficiency of BIPV/T-ASHP integrated heating system was relatively high in low temperature environment; |

### 3.1 ST-ASHP system

#### 3.1.1 R&D progress in simulation

From above table 2, in terms of ST-ASHP system, researchers using simulation aim to optimize the system, such as the structure, operation and efficiency, or to verify the performance of system, as some experiments cannot achieve in real projects. Weather data mainly containing temperature, solar radiation and seasonal change are the most important input parameters for simulation. The output parameters contain SPF, COP and tank volume and so on.

Most of the simulation studies focus on the optimization performance of the ST-ASHP system. Many researches compared their own proposed system with original ASHP system. For example, Jonas et al [63] developed and validated a tool--SHP-SimFrame in TRNSYS to simulate the solar heat pump system (SHP), especially solar thermal heat pump. Kim et al [75] carried out three indirect solar assisted heat pump systems including the serial type solar assisted heat pump, solar assisted heat pump with hybrid solar collectors and parallel solar assisted heat pump system with flat plate solar thermal collectors. Zhang et al [76] studied the structural parameters on the performance of a direct expansion solar assisted heat pump including the solar collector area, the collector thickness and the pipe length and internal diameter of condenser. And they got the optimized structural parameters for their system.

To make the table 2 clearer, we chose several researches including the special studies and a few typical studies. Almost all the simulation papers through software are conducted in TRNSYS software. i.e. Chen et al [64] used TRNSYS to optimize the ST-ASHP using $CO_2$ as heat pump refrigerant. The simulated results showed the optimized system has a high performance with the solar fraction can reach 71.1%, except Jonas et al [63], they developed and validated a tool 'SHP-SimFrame' in TRNSYS to simulate the solar heat pump system (SHP), especially solar thermal heat pump. They also only used the simulation method to compare the performance of the ST-ASHP system against the conventional ASHP system under various boundary conditions, but the conclusion was not representative, as the simulation parameters changed, the conclusion would also change.

Besides the simulation software method, Li et al [66] and Deng et al [67] developed their own mathematic models to simulate performance of ST-ASHP systems, respectively. Both of them calculated the COP value and found that solar irradiation and ambient air temperature have a great effect on the system performance. They indicated that the solar efficiency decreases with the growth of solar irradiation and decreases of ambient temperature. For detail information, Li et al [66] set up the parameters for components in TRNSYS and calculated SPF and SF of the system. It is proved that the overall efficiency of a DHW system was improved by the combination of a solar collector and heat pump. Deng et al [67] proposed a modified direct expansion solar assisted air source heat pump water heater system, which is illustrated in in Fig. 3. The proposed system mainly contained a flat-plate solar collector, an evaporator, a compressor, a hot water tank with heat exchanger as a condenser, receiver and two electronic expansion valves (EEV). There were two operation modes. When the solar radiant was high enough to heat the water efficiently, it run on single solar collector mode and. While during the solar radiant was unavailable or low, and evaporating temperature was much lower than ambient air, it worked under combination mode. As a result, at solar radiation of 100 $W/m^2$, the heating time of this system decreased by 19.8% compared to renovation when water temperature reaches 55 ℃. Meanwhile, the average system COP increased by 14.1%.

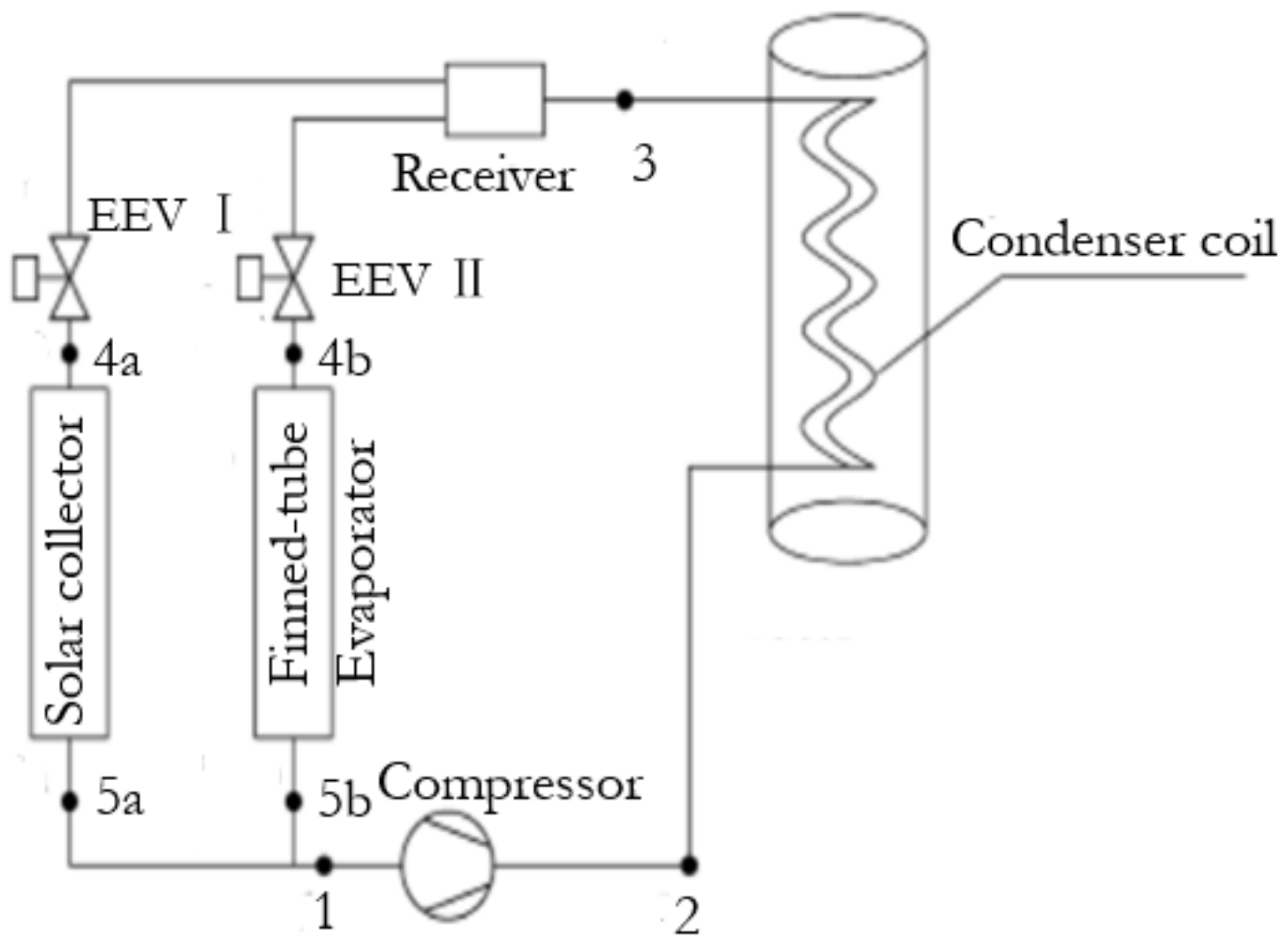

Fig. 3 Schematic diagram of the new system [67]

While all the results got by simulation were optimized. The performance of these studies were simulated under their defined simulation conditions. Among them, Fraga et al [68] not only simulated the achievable value of SPF, but also calculated the potential SPF value. They developed a simulation model to analyse the potential performance of a ST-ASHP system. SPF of system achieved 4.4 and when temperature distributed the SPF was 3.1-4.1. Moreover, SPF of system achieves five is potential, but considering the envelope, SH distribution temperature and solar collector area. In real project, the system should take account of real conditions e.g. investment and roof area.

### 3.1.2 ST-ASHP under extreme conditions

Among all the researches of solar assisted ASHP system, there is one special condition, that the performance researches under extreme conditions. The common problem under extreme conditions is that ASHP would be frosting, which also gained many concentrations in recent decades [77, 78]. There are few researches focused on ST-ASHP system that operated under extreme condition, which were not included in above table.

Qiu et al [79] compared the integrated heating system of solar energy and air source heat pump system under different working conditions in cold regions. As a result, compared with high temperature and low temperature heating collecting systems, medium temperature had the best performance of this integrated system. Its COP under medium temperature is 55% higher than other two types when the outdoor temperature is -25°C.

### 3.2 PV-ASHP system

For PV-ASHP, it was similar that most authors compared the energetic performance of their proposed system with original system or other system. For example, Evangelos et al [69] compared the ASHP with PV modules and the WSHP with flat plate collectors (FPC) and PVT system. They compared between ASHP with PV and PV/T for space heating. Giuseppe et al [87] set four simulations to compare the energy performance of the ASHP with PV modules system and conventional ASHP system by using the TRNSYS and Matlab. They assessed the performance in a detached house located in northern Italy with two different cases: one with a PV nodule and the conventional tile without PV. Their work was the application of a title for under-slating ventilation ducts, which was modelled in TRNSYS. They found that solar and air source combined heat pump system demonstrates as the most advantageous case compared to conventional ASHP systems.

While for economic performance, the related studies is few. Evangelos et al [69] examined an ASHP system with PV modules for space heating proposes. Fig.4 depicts the system concept. The electricity demand of the heat pump is covered partially by PV panels and partially by grid electricity import. To convert and regulate the voltage and store the momentary supplementary energy, an inverter and batteries are used in this system. They developed the related TRNSYS models and compared the performance, the electricity consumption and financial performance among four systems, which are PV-ASHP systems, water source heat pump with flat plate collectors (FPC-WSHP), water source heat pump with thermal photovoltaic collectors (PV/T-WSHP) and water source heat pump with photovoltaic and flat plate collectors (PV+FPC-WSHP). They concluded that when the electricity cost is between 0.2 €/kWh and 0.23 €/kWh, the use of 20 $m^2$ PV area with an air source heat pump is the most attractive solution financially in Greece.

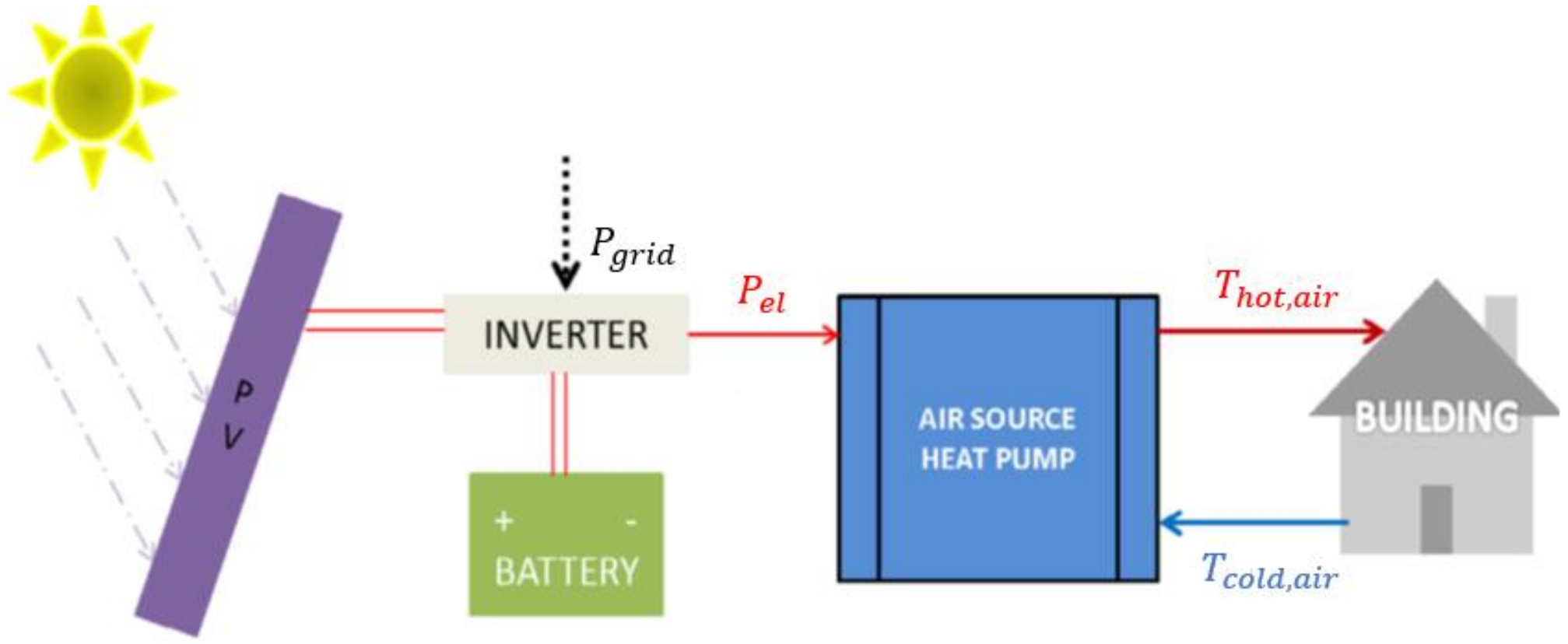

Fig.4 Air source heat pump heating system coupled with PV modules [69]

### 3.3 PV/T-ASHP system

Similar with all ASHP multi-system, the simulation tool for PV/T-ASHP system is also TRNSYS. And the performance of PV/T-ASHP system were the main studies contents, the main methods were to changed the parameters including the ambient condition or operation modes. They also proposed that, there exist technical problems to hinder the PV/T collector to be large scale used. For example, Raghad et al [71,72] proved in their TRNSYS model that an integrated BIPV/T+ASHP thermal energy system increases the overall COP and improves the energy efficiency of buildings.

Besides TRNSYS simulation, Cao et al [73] used the CFD simulation to test the performance of the PV/T-ASHP, which mainly studies the outlet temperature of the collector and the COP of the ASHP operated in severe cold region. Simulation time step was set to be 0.25h, tolerance integration and convergence were both 0.01. The control scheme was based on the load profile for the system. They obtained the result for supporting heating that the outlet temperature of PV/T collector can reach 76.6℃ and the COP of ASHP unit reached 4.1, which increased even in the low outdoor temperature.

Different from other studies that used simulation software, Getu et al [70] did the theoretical investigation of the performance of a two-stage variable capacity air source heat pump coupled with a BIPV/T system. They used the simulation method, and the BIPV/T system model was a TRNSYS type 568 component. The main results they found that the results were different along with ambient temperature. The COP of the new system increased obviously when the average temperature was between -3℃to 10℃; otherwise, there was no significant change.

## 4 Experiment methodologies and the related results

Table 3 summarizes the detailed experimental methods and the related results of solar assist ASHP systems. In this part, we listed the medium used in the experiment studies, the operation modes of the experiments, the practical system performance, the measured parameters and finally the main experiment results.

The most common medium for experiment researches of solar assisted ASHP systems are R410A, R407C and R134 a. R22 is also used in few papers. Unlike simulation, the operation conditions are only based on operation mode for experiment researches. Operation mode can be divided into heating mode, cooling mode and DHW mode.

Table 3 Summary of experiment methods and the related results of solar assisted ASHP systems

| Ref | System | Medium | Operation Mode | Location | Measured Parameters | Building Types | Evaluation Parameters | Main Results |
|---|---|---|---|---|---|---|---|---|
| [80] | ST- ASHP | R407c | heating mode | Taiyuan, China | environment condition | experiment room | exergy transfer;<br>COP;<br>energy and exergy efficiency;<br>process quality number;<br>improvement potential; | ▪Average exergy efficiency of SIASHP is 77.67, 8% higher than ASHP and the average COP is 2.94, 8.1% more than ASHP;<br>▪Average process quality number is 59.36, 5.4% higher than ASHP; |
| [81] | ST-ASHP | R407c | heating mode | Taiyuan, China | condensing temperature | experiment room | integrated part load value (IPLV);<br>seasonal part load value (SPLV);<br>relative size IPLV and SPLV; | ▪The higher part load rate, the higher COP;<br>▪SPLV is no more than IPLV;<br>▪Average IPLV of SIASHP is 2.54 and SPLV is 2.53, higher 14.9% and 15.5% than ASHP, respectively |
| [64] | solar combined $CO_2$ heat pump | — | solar heating/ $CO_2$ heat pump heating mode | Shanghai, China | ambient conditions | residential building | COP;<br>total heat gain;<br>solar fraction (SC) and SC efficiency; | ▪System COPs of the solar heating and $CO_2$ HP heating modes can reach 13.5 and 2.18;<br>▪The studied solar assisted system can save 53.6% of the electricity consumption than $CO_2$ HP; |
| [82] | Serial SA-ASHP | — | — | Erzurum, Turkey | weather data | residential building | COP;<br>environmental and economic benefits;<br>energy saving ratio;<br>$CO_2$ reduction ratio;<br>payback period; | ▪The average COP of heat pump is 3.8 and solar assisted ASHP is 2.9; |
| [83] | DMHP | — | heating /cooling mode | -- | ambient condition, temperature/ pressure of refrigerant, | enthalpy difference lab | power consumption and COP; exergy analysis; heating capacity and power consumption; water temperature in SWT and DWT | ▪ the COP is significantly affected by ambient temperature; ▪ Heating capacity and COP of air source space heating mode are higher than that of solar space heating mode with the outdoor ambient temperature above 4° C; ▪Different components cause most exergy loss in different operation mode |
| [68] [84] | ST-ASHP | — | space heating/ DHW | Geneva, Switzerland | weather data | Multifamily house | SPF; COP; energy flows; thermal storage; | ▪The measured seasonal performance factor of the system is 2.9 for 2012; ▪Several points may have contributed to a relatively low SPF: (i) an unusually low demand for space heating along with an unusually high demand for domestic hot water; (ii) in absence of a load adjusted heat pump; (iii) a single heat distribution circuit with decentralized domestic hot water storage; (iv) no insulation of the unglazed solar collectors |
| [85] | PV-ASHP | — | heating/ cooling mode | Changsha, central south China | different PV power | office building in university | exergy efficiency/consumption; life expectancy; $CO_2$ emission reduction; | ▪life expectancy of PV-ASHP system in Central south china is about 26 years; ▪Install PV capacity decide by the ASHP required rating power; ▪PV-ASHP has an exergy consumption saving rate of 41.16% for cooling and 35.02% for heating |
| [86] | PV-AC | R134a | heating mode | Alicante, Spain | irradiation; environment temperature; air mass flow rate; | residential buildings | heat recovered; heat production; COP; energy performance index; | ▪Average COP of heat pump increases from 3.6 to 3.75; ▪PV cell integrated in the tile can produce more energy than air conditioning required; |
| | | | | | | | primary energy consumption; | |
| [87] | PV module traditional HP system | — | Real operation condition vs. a scaled-down channel (1:5) | Northern Italy | the integral absorptivity and emissivity; the airflow pressure drop | residential building | Air temperature; Energy consumption; Integration angle; COP | ▪ air temperature variation ranges from 2 °C to 20 °C leading to a heat recovered between 2 kW and 7 kW in winter and summer time, respectively.<br>▪ the integration of the PV cell in the tile reduces the thermal energy by 15%.<br>▪The heat recovered enhances the heat pump COP, reducing the Primary Energy Consumption of a conventional heat pump by 5%, and achieved with limited additional costs. |
| [73] | PV/T-ASHP | R410 A | cooling mode | Shenyang, China | weather, thermal load, solar irradiance | residential building | power consumption; energy consumption; solar contribution; production factor; load factor; useful thermal energy; EER; | ▪System performance depended on the environment including solar radiation, outlet temperature and the unit's load factor;<br>▪Several technological improvements in the analyzed system have been made to optimize the system performance; |
| [72] | PV/T-ASHP with TES | — | exergy efficiency/exergy consumption cost model | Toronto, Canada | — | Test hut | model relative error; | ▪The relative error of exergy efficiency mode of integrated PV-ASHP for cooling is less than 4.21%;<br>▪The relative error of exergy consumption cost model is 1.5% for cooling and 0.3% for heating; |
| [88] | PV/T-ASHP | — | hot water | Beijing, China | ambient temperature and solar irradiance | laboratory | COP; | ▪COP of heat pump reduced from 5.61 to 1.69 and the average was 3.03.<br>▪Comprehensive system COP ranged from 6.07 to 1.33, the average was 2.99 |
| [89] [90] | ST-ASHP combined with PCM | — | heating/cooling/water from freezing | Shenyang, China | ambient temperature; solar hot water temperature; | laboratory | COP; Capacity; | ▪Ambient temperature had a significant influence on the system performance in cooling mode; ▪In heat pump cycle, solar-heated water temperature had a significant effect on system performance; |

Similarly, the outdoor conditions are also essential to the experimental studies, such as temperature and solar irradiation. The most measured parameters include temperature/pressure of compressor/condenser/refrigerant, input power and flow rate. The evaluation parameters are the same as that in simulation, including COP, SPF, power consumption and payback time.

Solar energy assisted ASHP system can save energy and benefit environment using sustainable solar energy [91,92]. Comparing to the traditional ASHP system, the average COP of solar assisted ASHP systems in experiments can increased from about 1 to almost 3.75. In the following subsection, we will extend the discussions specifically by selecting several researches of each solar assisted ASHP.

## 4.1 ST-ASHP system

### 4.1.1 R&D progress in experiment

Solar thermal assisted ASHP system can save energy and benefit environment using sustainable solar energy [54][93]. The main evaluation parameters studies by most researches include COP and energy relative parameters such as energy efficiency, energy consumption and energy saving ratio and so on, i.e., Wu et al [94] carried out the experiment of low-temperature ST-ASHP system for space heating. The solar thermal collectors were connected in parallel with the evaporator. The system has a mean COP of 3.2, when it is designed to supply heating to building in transition season and coldest winter days when the ASHP system cannot meet the demand. Only few papers also contain environmental and economic benefits. For example, Kadir et al [82] calculated the payback time of the solar thermal assisted ASHP according to the LPG, electric and fuel oil are 1.4, 2.9 and 3.9 year, respectively.

These studies usually conducted under different conditions, the typical study is Sun et al [94] and Xu et al [80, 81]. Sun et al [94] set experiment to compare the performance of the ASHP and solar assisted ASHP system for domestic hot water (DWH) in different outdoor condition and then simulated two systems for the whole year performance in Shanghai, China. They measured outlet temperature of two different systems in clear and overcast day and night and simulated the month average COP according to the ambient temperature over cast/rainy/snowy days. They concluded that the COP of the ST-ASHP system performs better especially in the winter, when the ambient temperature is low, since the ST-ASHP system draw the heat from the solar can increase the evaporating temperature more remarkably. Xu et al [80,81] proposed the solar integrated air source heat pump (SIASHP) with R407c. The system is mainly composed of the SIASHP and capillary copper pipe network without any other large component. They completed the experiment in the laboratory and the first generation of prototype of SIASHP was shown in Fig.5. In the system, the R407c absorbs the heat from both solar radiation and air source at the same time in the solar finned tube evaporator and runs through the compressor into the capillary copper pipe network, and then goes by the electronic expansion valve back into the solar finned tube evaporator to complete a whole working cycle. The solar finned tube evaporator bypass regulates the internal refrigerant flow of solar finned tube evaporator to avoid the excessive overheating of refrigerant into the compressor protecting the safe operation of compressor. The compressor bypass is the standard configuration for the direct current frequency conversion compressor to protect it from liquid hammer when underload. They claimed the average COP of SIASHP with the part load rate of 100% is 2.77, which is 9.8% higher than the 2.52 COP of ASHP and the SIASHP has a dramatically better heating performance with a low level of part load rate in the lab and analysed the energy efficiency and exergy efficiency of the SIASHP system. Taiyuan and a 130 $W/m^2$ solar irradiance, the 76.8% exergy efficiency of SIASHP is 7.9% higher than the 71.1% exergy efficiency of ASHP. From 9:00 to 21:00, the 2.89 kWh power consumption of SIASHP is 7.4% lower than the 3.12 kWh of ASHP.

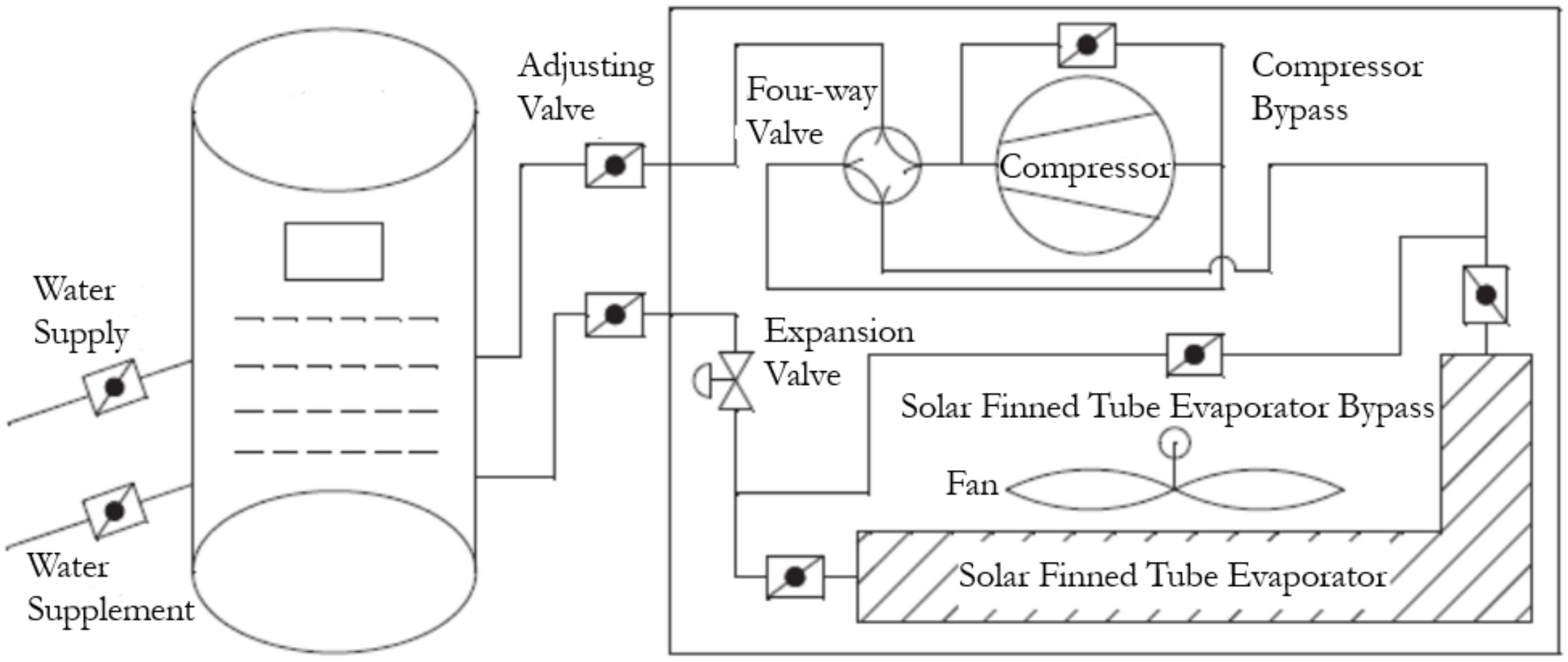

Fig.5 Schematic diagram of the SIASHP system [80,81]

While for mathematic analyse researches, most of them are under stable condition, which is easy to calculate. However, Cai et al [96] built mathematical model and done the thermodynamic analysis. they tested the performance of the IX-SAMHP system by using numerical analysis and compared the experiment results with simulated model in the space heating mode and space water mode condition. Cai [83] and Ji et al [95] proposed a novel indirect expansion solar-assisted multi-functional heat pump which composes of the multi-functional heat pump and solar thermal collecting system in Hefei, China. The schematic is shown in Fig.6. The system can fulfil space heating, space cooling and water heating with high energy efficiency by utilizing solar energy. The refrigerant circulation loop for the solar space heating is 1-2-9-7-6-5-2-3-1 and for the space cooling plus water heating mode is 1-2-5-6-8-9-3-2-1, while the water circulation loops are 12-14-13-12/13-11-5-13 and 5-10-11-5, respectively. Cai et al [96] investigate experimentally and theoretically on a novel dual source multi-functional heat pump (DMHP) system. The DMHP system can supply air conditioning and domestic water with air source or solar energy in different working modes. The DMHP system contains indoor units and outdoor units linked by refrigerant circuit and water circuit. The refrigerant circuit is a multi-functional heat pump system consisting of two air heat exchangers, a plate-type heat exchanger, a domestic water tank and a compressor. They tested the influence factor which affect the system including the initial solar water tank, the solar irradiation in the solar water heating mode and solar space heating mode. For the space heating mode, the increase of the indoor environment temperature decreases the heating capacity and COP. Increasing the initial water temperature in solar water tank can improve the condensing power and evaporation power, as well as the energy consumption and COP of the system. Moreover, the higher solar irradiation can lead to higher heat transfer rate and larger energy consumption. In annual analysis, the DMHP system can obtain relatively high COP of the value above 2.0 throughout the year with the optimal working strategy in three cities under different climate conditions.

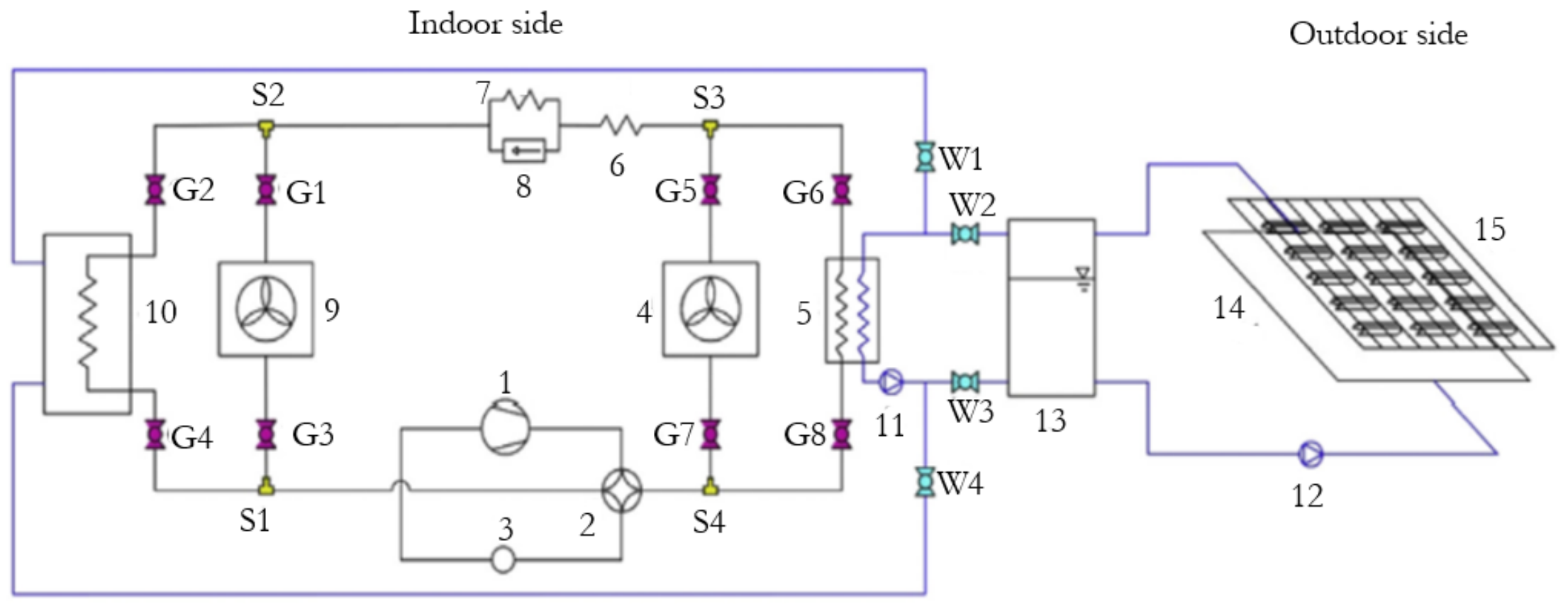

1.compressor 2.reversing valve 3.liquid accumulator 4.outdoor air heat exchanger 5.plate-type heat exchanger 6~7.caillary tube 8.one-way valve 9.indoor air heat exchanger 10.domestic water tank 11~12.water pump 13.solar water tank 14.solar flat-plat collector 15.solar simulator G1~G8 refrigeration valve S1~S4 three-way valve W1~W4 water valve

Fig.6 Schematic of the solar-assisted multi-functional heat pump system [83,95]

One special study among them is that Chen et al [64]. They investigated a solar combi-system consisting of solar collector and a CO2 heat pump to analyse its performance. Chen et al [6464] conducted the experiment study on the performance of a pre-existing solar combi-$CO_2$ heat pump system. Fig.7 shows a schematic diagram of the system in residential building. The system separates the operation tank from the storage tank. The storage tank acts as the main storage of the captured solar energy and when the temperature level achieved, it delivers energy to the operation tank. The $CO_2$ heat pump will consume less electricity to maintain the temperature in the small volume operation tank when solar radiation is poor. The operation tank can also be a buffer of heat for whole system with the obvious temperature changes caused by solar radiant. It can weaken the temperature fluctuation. The result showed that the solar assisted system can save electricity consumption at 1790.8 kWh every year with an average COP of 13.5, comparing with CO2 HP heating system based on year-round operation (COP is 2.18).

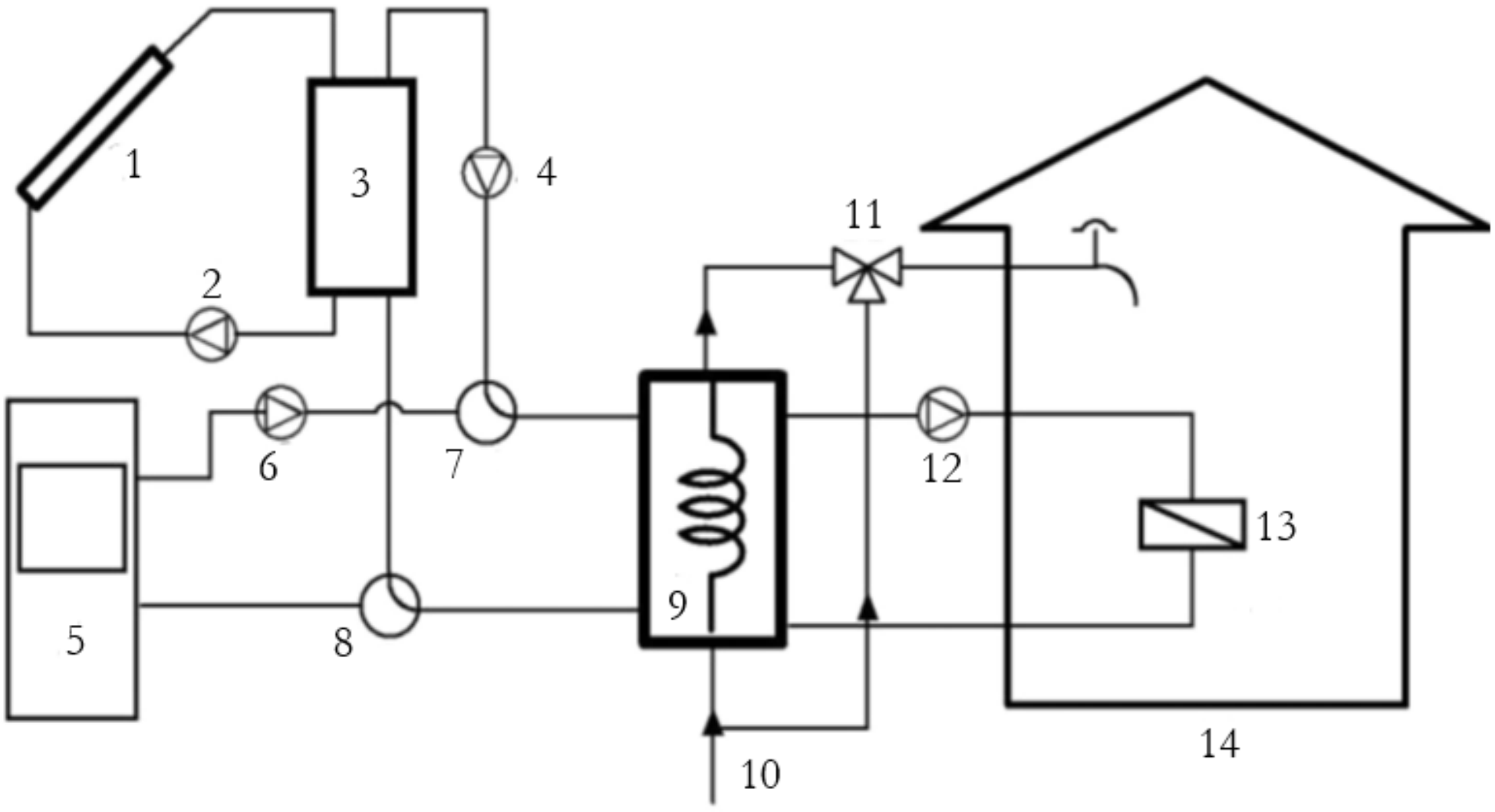

1.Solar collector array 2. solar collector pump 3. storage tank 4. solar collector delivery pump 5.$CO_2$ heat pump 6.heat pump delivery pump 7.8.switching valve 9.operation tank 10. civil water inlet 11.mixing valve 12.supply pump 13.fan coil units 14.testing room

Fig.7 Diagram of solar combi-system with $CO_2$ HP [64]

### 4.1.2 ST-ASHP under extreme conditions

For extreme condition studies on the ST-ASHP system, there are only few papers conducted the related research in recent years. The related research mainly concentrated on the performance of the system by controlling different operation modes or under different ambient conditions [97]. Huang et al [98, 99] experimentally investigated the frosting characteristics of solar collectors with direct expansion air source heat pump system. They conducted the experiment by supporting different temperature, controlling the air humidity and solar irradiation in a lab. As a result, the ambient condition including the temperature, humidity and solar irradiation had a significant on the performance of this system under frosting conditions. The frost occurred when the ambient humidity was 50% to 70%, the temperature ranged from 7 ºC to 6 ºC with zero solar irradiation. While the solar irradiation was 100 W/m2, there was no frosting except the ambient temperature was lower than -3 ºC and the humidity is higher than 90%. Long et al [100] studied the performance of a dual-solar and sir heat source integrated heat pump evaporator in cold season. The hot water of evaporator had an obvious effect on refrigerant's evaporator temperature and the COP of the heat pump. Liu et al [101] found that the heat capacity and COP of solar assisted air heat pump system increased 62% and 59%, compared with air source heat pump system, when the ambient temperature was -15 ºC.

### 4.2 PV-ASHP system

The common researches for PV-ASHP system also concentrated on the energetic performance, i.e., Giampaolo et al [87] used the experimental to assess the performance of a PV-ASHP system and validate their simulation model. They adopted a similar approach as in [102] to determine the flow rate and temperature of the air at the outlet of the ventilation channel and combined the generalized They found that the average COP increased from 3.60 to 3.75, turning into a primary energy reduction with respect to the reference case of 5%.

Besides the energetic performance, some studies also evaluated the environmental and economic performance. Wang et al [85] set up an experiment of the integrated PV-ASHP system and tested the system in six cases with different PV powers. Their integrated PV-ASHP system was made up of three sub-systems: a PV system, an electricity storage, an inverter system and an ASHP system, and the experiment located on the top floor of a building. They indicated that the PV-ASHP system could have a saving rate of 41.16% of exergy consumption per unit investment for cooling and 35.02% for heating. The life expectancy of the PV-ASHP system could reach about 26 years and it can reduce the 11.10t of $CO_2$ in the whole life emission and save the operation cost comparing with the ASHP powered by electricity from the national grid.

Besides the conventional PV/T-ASHP system, Aguilar et al [86] conducted an experiment using photovoltaic air conditioning unit (PV+AC). They carried out the experiment on an air conditioning unit which has been powered using both a PV installation and the grid simultaneously for a whole year with the control system, which is designed to maximize solar contribution in Spain. The PV+AC system consisted of the air conditioning unit with two electrical connections (PV panels and grid) and the PV installation. They test the PV+AC system from May to October in cooling mode in an office located in Alicante (Spain) and the whole system has demonstrated to be 100% reliable, having undergone no maintenance. Then they optimized the operation mode, regulating the air conditioning unit independently its own operation regime to maximize the PV energy input.  The solar contribution obtained in cooling mode from May to October was 64.5%, while the production factor was 65.1%. The performance of the system will depend on the solar radiation, outlet temperature and the unit's load factor.

### 4.3 PV/T-ASHP system

For PV/T-ASHP system, few papers done the experiment research, many authors just pay attention to the simulation as the system is complex and difficult to set up in laboratory. And most of the researches focused on the energetic performance. For example, Wang et al [88] built experiment of the PV/T-ASHP hot water system in laboratory and tested the system. The system was comprised of independently developed flat plate solar PV/T collector based on micro-channel heat pipe array and air source heat pump, which were combined by a new composite evaporator. The COP of heat pump reduced from 5.61 to 1.69 and the average was 3.03. Comprehensive COPsys of PV/T-air composite heat pump system ranged from 6.07 to 1.33, the average was 2.99. Raghad et al [64] design BIPV/T-ASHP system on a test hub and found that the seasonal COP could be increased from 2.74 to a maximum value of 3.45 for direct coupling of BIPV/T+ASHP without the use of diurnal thermal storage. The heat pump electricity consumption is reduced by 20% for winter. The PV/T-ASHP system usually composed by three parts: (1) PV/T collector or BIPV/T system on facade; (2) heat pump unit; and (3) thermal energy storage (TES) system, including Insulated Concrete Forms walls, ventilated concrete slab, gravel/sand bed and water tank storage. Qu et al [103] examined a novel solar photovoltaic/thermal integrated dual source heat pump water heating system at different ambient temperature. Besagni et al [104] focused on the solar-assisted heat pumps for heating and cooling to produce DHW. The system was coupled with PV or PV/T system. They made the examination under different ambient temperature, system parameters and operating modes to compare the performance of the system, main the COP.

One special condition facing by ASHP system in extreme condition is frosting. To avoid the deterioration of heat transfer effect and performance drop of unit due to the evaporator frosting in low temperature condition with the circulated air heated by the air collector, Cao et al [73] and Li et al [74] proposed a PV/T-ASHP integration heating system, which aimed to operate safely, stably and efficiently in winter in cold region. PV/T air collectors were arrayed on the building envelop composing an integrated PV/T curtain wall. In the PV/T, up and down sides are respectively opened outlets and inlets connected to the ASHP unit with air ducts forming a forced circulation in the system. Besides, they assumed that there is almost no overheating during the energy transfer process, freezing, boiling, corrosion and leakage problems. Lu et al [105] studied the performance of PV/T-AHP system in winter. To improve the performance, they added vapor injection to this system. As a result, the COP was 3.45 when the ambient temperature average was -1.13 °C and solar irradiation was 164.03 W/m$^2$.

**5 Comparison of different solar assisted ASHP systems**

For conventional ASHP system, compared with other source heat pump system, such as, water, ground and so on, the main advantages of ASHP system conclude that the investment is much lower and the installation is much easier than other ground source heat pump system, and the operation and management of ASHP system also easier. However, the disadvantages are also obvious, the ASHP system has a more serious for ambient environment. Water and ground source heat pump systems are more stable and the performance of them are higher than ASHP system, especially under extreme conditions, the frosting is still one of big problems for ASHP system, which seriously affects the performance of ASHP. Auxiliary heat will increase the extra electricity energy consumption.

For solar assisted air source heat pump system, including the ST-ASHP, PV-ASHP and PV/T-ASHP system, compared with ASHP system, the common advantages for them are obvious. From the system aspect, performance of them are better than ASHP system [106], as the solar assisted systems gain energy from both air and solar, which are both renewable and sustainable. In before reviewed part, the solar assisted air source heat pump system could increase the COP about 30% to 60% than ASHP system, under some defined condition, the value is higher. The frosting problem also reduced as the solar supply heat for the ASHP system, as a result, the solar assisted air source heat pump system has a wider spread than ASHP system, special for cold and more extreme regions [21]. Moreover, for the whole energy aspect, the solar assisted air source heat pump system could supply more energy to users and produce more types of energy, i.e., PV-ASHP and PV/T-ASHP system can supply electricity for buildings. Solar assisted air source heat pump system could meet the increasing energy demand better than ASHP system. Besides that, as mentioned above, the solar and air energy are clean and friendly to environment, the solar assisted systems could reduce the environment pollution caused by fossil fuels, i.e. coil, oil. Along these obvious advantages, it is certainly that there are some common weaknesses for all solar assisted air source heat pump systems. Comparing with ASHP system, the investment concluding equipment, installation and maintenance is much higher and the operation and management are much more complex, as the systems are complex.

While only for solar assisted air source heat pump systems, ST-ASHP, PV-ASHP and PV/T-ASHP system, there are also some comparison results of their benefits. Among solar assisted air source heat pump systems, ST-ASHP can avoid frosting in winter especially in cold regions. Even below ambient temperature, the system performance is still relatedly high. Moreover, the installation requirement is much less strict than PV-ASHP and PV/T-ASHP systems, causing that the payback time is the shortest among three solar assisted air source heat pump system. However, the disadvantage is that most energy produced by ST-ASHP is used only for heat or hot water, and ST-ASHP system is still affected more by the external weather conditions than other two systems, as ST-ASHP system is significantly related with ambient temperature, while other two solar assisted air source heat pump systems mainly rely on the solar radiation. In terms of PV-ASHP system, this system is usually parallel, the PV cell produce electricity supplied to equipment consumption or directly to buildings, while ASHP system supply the heat and hot water (with water tank system) to users. Besides the heat energy, electricity energy can be produced for direct use, which is improved than ST-ASHP system. The performance of the air source heat pump is the best among three solar assisted air source heat pump system, as the energy input is much lower. Of course, more complex system control strategy than ST-ASHP system is also required among PV generation, grid import, battery or TES unit. And electricity storage is still one of difficulties which hinder the widespread use. As for the most complex system among solar assisted air source heat pump systems, PV/T-ASHP system has its own characteristics. It can produce different types of energy including heat, hot water and electricity. Similar with PV-ASHP, but not the same, it can produce more capacity of energy and the users consume the lowest energy among three solar assisted air source heat pump systems, as this system has the highest and maximized solar utilization. But at this step, there are also some problems facing by PV/T-ASHP system. The most complex control system is basic for operation and management. The maintenance is much difficult. All these cause that it has the longest payback time. These characteristics restrict the development of PV/T-ASHP systems [44,107]. The summary of the main characters of solar assisted air source heat pump systems is given in Table 4.

Table 4 Summary of comparison among solar energy assisted ASHP systems

| System | Advantage | Disadvantage | Average COP of HP |
|---|---|---|---|
| ST-ASHP | High efficiency;<br>Shortest payback time;<br>Low installation requirement; | Affected by ambient temperature;<br>lowest investment; | 2.9 |
| PV-ASHP | Saving electricity, even can support other equipment;<br>Highest COP of HP;<br>Lower input power; | More investment;<br>Existing electricity storage problem;<br>Affected by solar irradiation;<br>Installation problem; | 3.75 |
| PV/T-ASHP | Highest solar energy utilization;<br>Supply heat and electricity as the same time;<br>Most energy production and lowest consumption; | Most investment;<br>Most complex system and control;<br>Most installation and maintenance problem;<br>Longest payback time; | 3.03 |

## 6 Limitations and future directions

There are still some limitations which need to be broken through before the solar assisted ASHP system can be applied widespread in market. The main issues are concluded in the following.

Methodologies adopted in existing literatures varies from researcher to researcher. Each one can set up different experiment mode, define boundary conditions and choose different models. Such inconsistence limits the marketization promotion of solar assisted ASHP systems. Another challenge lies in gap between solely simulation or laboratory testing and real application [67]. In the future, there are strong needs to: (1) develop a common simulation tool; (2) measure the system performance under a standard testing conditions defined by testing science, and (3) collect field testing data from pilot projects and validate the simulation/experiment results.

The most studied system is ST-ASHP type. On the contrary, experiment of (BI)PV/T-ASHP is covered by only few papers, and the rest papers on this system concept are all from simulation points of view. This means that the (BI)PV/T-ASHP has not yet been largely applied or there are still practical problems, such as technic or installation, to be solved before it can get wider application. In addition, optimization of the systems is required, including the structure optimization of some component in the whole system [80,81], the operation mode optimization as their running mode is adjusted by time which is not accurate and the capacity or volume optimization about different components [65].

There is lack of standardized indictor to evaluate the system performance [108,109]. Although COP and SPF are adopted in most papers, the process and the evaluation parameters are different from each to the other. Moreover, few papers [66, 82] take into account of the environmental and economic parameters. Common standards and official certifications can help users to make better choice and comparison among different solar assisted ASHP systems.

While solar assisted ASHP system suffers from existing limitations in single system, combined two or more system may have a prospective development in the future, e.g. combination of the heat recovery system and solar ASHP system, which can avoid the disadvantages in single system and provide more solutions to buildings energy systems.

## 7 Conclusions

The aim of this paper is to conduct a systematic review about three main solar assisted ASHP systems. In order to give useful information for users on how to choose and implement appropriate system in practice, a detailed description of each system is presented, including their configurations, working principle, research methods, main performance and advantages/disadvantages. However, there are many other aspects, such as economics, environmental issues, which need to be discussed in the future. Conclusions are generally summarized as below:

It is so important to define the boundary conditions of solar assisted ASHP system from the beginning, according to the IEA SHC Task 44/HPP Annex 38. This can help to evaluate the system performance in a standard way. The experiment is usually conducted in laboratory and simulation is conducted by TRNSYS software. COP is significant to evaluate the performance of the system and adopted in most papers. Weather data in different locations is mostly sensitive to the system performance, which contains temperature and solar radiation etc.

Among the three systems, the ST-ASHP system is mostly studied, which has relatively low COP at mean value of 2.9, but requires much smaller investment with 3.5-10 years as payback time. Either PV-ASHP or PV/T ASHP has a bit higher COP value than that of ST-ASHP, at 3.75 and 3.03 respectively, which, however, needs larger capital cost (longer payback time) and more complex control strategy.

Further researches are recommended from three aspects:

- **Methodologies**: (1) develop a common simulation tool; (2) measure the system performance under a standard testing conditions, and (3) collect field testing data from pilot projects and validate the simulation/experiment results.
- **Systems**: (1) (BI)PV/T-ASHP system type needs largely developed; (2) optimization of the systems is required, including structure, control and operation to improve performance of the system; (3) combine two or more system concepts may have a prospective development, e.g. combination of the heat recovery system and solar ASHP system, which can avoid the disadvantages in single system and provide more solutions to buildings energy systems.
- **Standard indicators**: common standards and official certifications can help users to make better choice and comparison among different solar assisted ASHP systems.

This review is expected to clarify the current R&D status of solar assisted ASHP systems and promote their development to achieve widespread application at building level.

## Acknowledgement

The authors acknowledge the financial support from the University of Nottingham Ningbo China, Beijing Natural Science Foundation of China (Grant No. 3172041) and the EU Horizontal 2020 Energy-Matching project (Grant No. 768766). The authors also want to acknowledge the useful information and materials from IEA SHC Task 44 and Task 60.

## References:

[1] T Makoto. Regional report Asia and Pacific. In: 9th international IEA heat pump conference, Zurich, Switzerland. 2008, May 20-22

[2] A Philippe, L Jean. EBC Annex 48: Heat pumping and reversible air conditioning-Project summary report. 2016,08,04

[3] L Jin, F Cao, D Yang, X Wang. Performance investigations of an R404A air source heat pump with an internal heat exchanger for residential heating in northern China. Int J Refrig 2016;67: 239–48

[4] Y.M. Chen, L.L. Lan, A fault detection technique for air-source heat pump water chiller/heaters, Energy Build. 41 (8) (2009) 881–887

[5] K. Gram-Hanssen, T.H. Christensen, P.E. Petersen. Air-to-air heat pumps in real-life use: are potential savings achieved or are they transformed into increased comfort, Energy Build. 53 (2012) 64–73

[6] Y. Jiang, T.S. Ge, R.Z. Wang, Y. Huang, Experimental investigation on a novel temperature and humidity independent control air conditioning system –part II: heating condition, Appl. Therm. Eng. 73 (1) (2014) 775–783

[7] M.J. Song, N. Mao, S.M. Deng, Y.D. Xia, Y. Chen. An experimental study on defrosting performance for an air source heat pump unit with a horizontally installed multi-circuit outdoor coil, Appl. Energy 165 (2016) 371–382

[8] G Ma, Q Chai, Y Jiang. Experimental investigation of air-source heat pump for cold regions. Int J Refrig 2003;26:12–8

[9] Y.Q. Jiang, J.K. Dong, M.L. Qu, S.M. Deng, Y. Yao, A novel defrosting control method based on the degree of refrigerant superheat for air source heat pumps, Int. J. Refrig. 36 (8) (2013) 2278–2288

[10] G.Y. Ma, Q.H. Chai, Y. Jiang, Experimental investigation of air-source heat pump for cold regions, Int. J. Refrig. 26 (1) (2003) 12–18

[11] F. Qin, Q.F. Xue, G.M.A. Velez, G.Y. Zhang, H.M. Zou, C.Q. Tian, Experimental investigation on heating performance of heat pump for electric vehicles at $-20^{\circ}$ C ambient temperature, Energy Convers. Manage. 102 (2015) 39–49

[12] S. Jiang, S. Wang, X. Jin, Y. Yu, Optimum compressor cylinder volume ratio for two-stage compression air source heat pump systems, Int. J. Refrig. 67 (2016)77–89

[13] Q Zhang, L Zhang, J Nie, Y Li. Techno-economic analysis of air source heat pump applied for space heating in northern china. Applied Energy 207 (2017) 533–542

[14] A Hesaraki, S Holmberg, F Haghighat. Seasonal thermal energy storage with heat pumps and low temperatures in building projects—A comparative review Renewable and Sustainable Energy Reviews 43(2015)1199–1213

[15] Z Wang, F Wang, X Wang, Z Ma, X Wu, M Song. Dynamic character investigation and optimization of a novel air-source heat pump system. Applied Thermal Engineering 111 (2017) 122–133

[16] LZ Zhang. Total heat recovery: heat and moisture recovery from ventilation air. New York: Nova Science Publishers, Inc; 2008.

[17] S.J. Sterling, M.R. Collins, Feasibility analysis of an indirect heat pump assisted solar domestic hot water system, Appl. Energy 93 (2012) 11–17

[18] S Mekhilef, A Safari, WES Mustaffa, R Saidur, R Omar, MAA Younis. Solar energy in Malaysia: current state and prospects. Renewable Sustainable Energy Rev 2012; 16:386–96

[19] R Kumar, MA Rosen. Acritical review of photovoltaic—thermal solar collectors for air heating. Appl Energy2011;88(11):3603–14

[20] YH Yau, WC Chan, CWF Yu. Solar thermal systems for large high rise buildings in Malaysia. Indoor Built Environ 2014; 23(7):917–923(7)

[21] BJ Huang, T Lin, WC Hung, FS Sun. Performance evaluation of solar photovoltaic/thermal systems. Sol Energy 2001;70(5):443–8

[22] G Xu, S Deng, X Zhang, L Yang, Y Zhang. Simulation of a photovoltaic/thermal heat pump system having a modified collector/evaporator. SolEnergy2009;83 (11):1967–76

[23] HB Nejma, A Guiavarch, I Lokhat, E Auzenet, F Claudon, B Peuportier. In-situ performance evaluation by simulation of a coupled air source heat pump/PV-T collector system. In: Proceedings of BS2013:13th conference of international building performance simulation association, Chambery, France; August 26– 28, 2013.p.1927–1935

[24] A Kylili, PA Fokaides. Investigation of building integrated photovoltaics potential in achieving the zero carbon building target. Indoor Built Environ 2014;23(1):92–106

[25] G Xydis. Exergy analysis in low carbon technologies—the case of renewable energy in building sector. Indoor Built Environ 2009;18(5):396–406

[26] Survey of Energy Resources 2007, World Energy Council.

[27] H.S. Rauschenbach. Solar Cell Array Design Handbook: The Principles and Technology of Photovoltaic Energy Conversion

[28] AR Jordehi. Parameter estimation of solar photovoltaic (PV) cells. Renew Sustain Energy Rev 2016; 61:354–71

[29] Burger, Meyer, 2013. Hybrid. The Intelligent Combination of Solar Thermal Energy and Photovoltaics. Available from. http://energysystems.meyerburger.com/en/products/hybrid/hybrid/ (last accessed 10.05.16)

[ 30 ] Minimise Group, 2015. Hybrid PV-T Range. Available from. (last accessed 12.05.16). http://www.minimisegroup.com/hybrid-solar-PV/T-panels.html

[ 31 ] Solar Wall, 2016. PV/Thermal; Solar Power Wall e Electricity t Heating. Available from. http://solarwall.com/en/products/PV/Thermal.php (last accessed10.05.16)

[32] V Delisle, M Kummert. Cost-benefit analysis of integrating BIPV-T air systems into energy-efficient homes. Solar Energy 136 (2016) 385–400

[33] JC Hadorn. Solar and heat pump systems for residential buildings. John Wiley &Sons; 2015

[34] IEA, SHC. Task 53. New Generation Solar Cooling & Heating Systems - Available o;2017. ⟨http://task53.iea-shc.org⟩ [-05-23]. n.d.

[35] P Stefano, S Nelson, B Chris, M Hatef, L Per. Techno-economic review of solar heat pump systems for residential heating Applications. Renewable and Sustainable Energy Reviews 81 (2018) 22–32

[36] E Frank, H Michel, S Herkel and R Jorn. Systematic classification of comined solar thermal and heat pump systems. Proceedings of the EuroSun 210 conference, Sep.29-Oct 1, Graz, Austria

[37] E Frank, MY Haller, S Herkel, J Ruschenburg, 2010. Systematic classification of combined solar thermal and heat pump systems. In: Proceedings of the EuroSun2010 Conference. Graz, Austria.

[38] J Ruschenburg, S Herkel, 2013. A review of market-available solar thermal heat pump systems. A technical report of subtask A. IEA SHC Task 44/HPP Annex 38.

[39] J Ruschenburg, S Herkel, HM Henning, 2013. A statistical analysis on market available solar thermal heat pump systems. Sol. Energy 95, 79–89

[40] Survey of Energy Resources 2007, World Energy Council

[41] A Joyce, L Coelho, J Martins, N Tavares, R Pereira, P Magalhães. A PV/T and Heat Pump Based Trigeneration System Model for Residential Applications. ISES – Sol. World Congr 2011 2011: p. 1–12.

[42] Ali H.A. Al-Waeli, K. Sopian, Hussein A. Kazemb, Miqdam T. Chaichan. Photovoltaic/Thermal (PV/T) systems: Status and future prospects. Renewable and Sustainable Energy Reviews 77 (2017) 109–130

[43] Zhongzhu Qiu, Xiaoli Ma, Xudong Zhao, Peng Li, Samira Ali. Experimental investigation of the energy performance of a novel Micro-encapsulated Phase Change Material (MPCM) slurry based PV/T system. Applied Energy 165 (2016) 260–271

[44] E Giuseppe, Z Angelo, DC Michele. A heat pump coupled with photovoltaic thermal hybrid solar collectors: A case study of a multi-source energy system. Energy Conversion and Management 151 (2017) 386–399

[45] L Cabrol and P Rowley. Towards Low Carbon Homes –A Simulation Analysis of Building-Integrated Air-Source Heat Pump Systems. Energy and Buildings http://dx.doi.org/10.1016/j.enbuild.2012.01.019

[46] T Christos, B Evangelos, M Georgios, A.A Kimon, D A simakis. Energetic and financial evaluation of a solar assisted heat pump heating system with other usual heating systems in Athens. Applied Thermal Engineering 106 (2016) 87-97

[47] B Evangelos, T Christos. Energetic and financial sustainability of solar assisted heat pump heating systems in Europe. Sustainable Cities and Society 33 (2017)70-84

[48] TP Maria, B Evangelos, T Christos, A.A Kimon. Financial and energetic evaluation of solar-assisted heat pump underfloor heating systems with phase change materials. Applied Thermal and Enginnering 149 (2019) 548-564

[49] P Stefano, B Chris, H Andreas, H Franz, C David et al.. Analysis of system improvements in solar thermal and air source heat pump combisystems. Applied Energy, Elsevier,2016, 173 (1), pp.606-623. <10.1016/j.apenergy.2016.04.048>. <cea-01310230>

[50] S Nikoofard, VI Ugursal, MI Beausoleil. Economic analysis of energy upgrades based on tolerable capital cost. J Energy Eng2014;141(3). http://dx.doi.org/10.1061/(ASCE)EY.1943-7897.0000203 [p. 06014002–1–6].

[51] P Khagendra. Bhandari, M Jennifer. J Collier, Randy. SA Ellingson. Energy payback time (EPBT) and energy return on energy invested (EROI) of solar photovoltaic systems: A systematic review and meta-analysis. Renewable and Sustainable Energy Reviews 47(2015)133–141

[52] M. Hosenuzzaman, N.A.Rahim, J.Selvaraj, M.Hasanuzzaman, A.B.M.A.Malek, A.Nahar. Global prospects, progress, policies, and environmental impact of solar photovoltaic power generation. Renewable and Sustainable Energy Reviews 41(2015)284–297

[53] Environmental management – life cycle assessment – principles and frame work;2006

[54] T.T. Chow A review on photovoltaic/thermal hybrid solar technology. Applied Energy 87 (2010) 365–379

[55] S. Rasoul Asaee, V. Ismet Ugursal, Ian Beausoleil-Morrison. Techno-economic assessment of solar assisted heat pump system retrofit in the Canadian housing stock. Applied Energy 190 (2017) 439–452.

[56] LG Swan, VI Ugursal, MI Beausoleil. Hybrid residential end-use energy and greenhouse gas emissions model – development and verification for Canada. J Build Perform Simul 2013;6(1):1–23. http://dx.doi.org/10.1080/19401493.2011.594906

[57] AA Farhat, VI Ugursal. Greenhouse gas emission intensity factors for marginal electricity generation in Canada. Int J Energy Res 2010;34(15):1309–27.

[58] https://en.wikipedia.org/wiki/TRNSYS

[59] MA Zakaria, MNA Hawlader. A review on solar assisted heat pump systems in Singapore. Renewable and Sustainable Energy Reviews 26 (2013) 286-293

[60] P Stefano and B Chris, "Techno-Economic Analysis of a Novel Solar Thermal and Air-Source Heat Pump System" (2016). International Refrigeration and Air Conditioning Conference. Paper 1638. http://docs.lib.purdue.edu/iracc/1638

[61] P Stefano, B Chris, Y Michel. Haller, Andreas Heinz. Influence of boundary conditions and component size on electricity demand in solar thermal and heat pump combisystems. Applied Energy 162 (2016) 1062–1073

[62] S Mahmut, S Riffat. Solar assisted heat pump systems for low temperature water heating applications: A systematic review. Renewable and Sustainable Energy Reviews 55(2016)399–413

[63] D Jonas, D Theis, F Felgner, and G Frey. A TRNSYS-Based Simulation Framework for the Analysis of Solar Thermal and Heat Pump Systems. ISSN 0003-701X, Applied Solar Energy, 2017, Vol. 53, No. 2, pp. 126–137

[64] J.F. Chen, Y.J. Dai, R.Z. Wang. Experimental and theoretical study on a solar assisted CO2 heat pump for space heating. Renewable Energy 89 (2016) 295-304

[65] H Li, Y Sun. Operational performance study on a photovoltaic loop heat pipe/solar assisted heat pump water heating system. Energy and Buildings 158 (2018) 861–872

[66] YH Li, WC Kao. Performance analysis and economic assessment of solar thermal and heat pump combisystems for subtropical and tropical region. Solar Energy 153 (2017) 301–316

[67] WS Deng, JL Yu. Simulation analysis on dynamic performance of a combined solar/air dual source heat pump water heater. Energy Conversion and Management 120 (2016) 378–387

[68] C Fraga, P Hollmuller, F Mermoud, B Lachal. Solar assisted heat pump system for multifamily buildings: Towards a seasonal performance factor of 5? Numerical sensitivity analysis based on a monitored case study. Solar Energy 146 (2017) 543–564

[69] E Bellos, C Tzivanidis, K Moschos, A Kimon. Antonopoulos Energetic and financial evaluation of solar assisted heat pump space heating systems Energy Conversion and Management 120 (2016) 306–319

[70] G Hailu, P Dash, AS Fung. Performance Evaluation of an Air Source Heat Pump Coupled with a Building-Integrated Photovoltaic/Thermal (BIPV/T) System under Cold Climatic Conditions. Energy Procedia 78 (2015) 1913 – 1918

[71] R Kamel, N Ekrami, P Dash, Alan Fung, Getu Hailu. BIPV/T+ASHP: Technologies for NZEBs Energy Procedia 78 (2015) 424 – 429

[ 72 ] RS. Kamel, S Alan. Fung. Modeling simulation and feasibility analysis of residential BIPV/T+ASHP system in cold climate—Canada. Energy and Buildings 82 (2014) 758–770

[73] C CAO, H Li, G Feng, R Zhang and K Huang. Research on PV/T-air source heat pump integrated heating system in severe cold region. Procedia Engineering 146 (2016) 410 – 414

[74] H Li, C Cao, G Feng, R Zhang, K Huang. A BIPV/T System Design Based on Simulation and its Application in Integrated Heating System. Procedia Engineering 121 (2015) 1590 – 1596

[75] T Kim, B Choi, YS Han, KH Do. A comparative investigation of solar-assisted heat pumps with solar thermal collectors for a hot water supply system. Energy Conversion and Management 172(2018)472-484

[76] D Zhang, QB Wu, JP Li, XQ Kong. Effects of refrigerant charge and structural parameters on the performance of a direct expansion solar assisted heat pump system. Applied Thermal Engineering 73 (2014)522-528

[77] MJ Song, L Xia, N Mao, SM Deng. An experimental study on even frosting performance of an air source heat pump unit with a multi-circuit outdoor coil. Applied Energy 164 (2016) 36-44

[78] MJ Song, N Mao, YJ Xu, SM Deng. Challenges in and the development of building energy saving techniques, illustrated with the example of an air source heat pump. Thermal Science and Engineering Progress 10 (2019) 337-356

[79] GD Qiu, XH Wei, ZF Xu, WH Cai. A novel integrated heating system odf soalr energy and air source heat pumps and its optimal working condition range in cold regions. Energy Conversion and Management 174 (2018) 922-931

[80] D Xu, Q Tian, Z Li. Energy and exergy analysis of solar integrated air source heat pump for radiant floor heating without water. Energy and Buildings 142 (2017) 128–138

[81] D Xu, Q Tian, Z Li. Experimental investigation on heating performance of solar integrated air source heat pump. Applied Thermal Engineering 123 (2017) 1013–1020

[82] K Bakirci, B Yuksel. Experimental thermal performance of a solar source heat-pump system for residential heating in cold climate region. Applied Thermal Engineering 31 (2011) 1508-1518

[83] J Cai, J Ji, Y Wang, W Huang. Numerical simulation and experimental validation of indirect expansion solar-assisted multi-functional heat pump. Renewable Energy 93 (2016) 280-290

[84] Fraga, C., Mermoud, F., Hollmuller, P., Pampaloni, E., Lachal, B., 2015. Large solar driven heat pump system for a multifamily building: long term in-situ monitoring. Solar Energy 114, 427–439

[85] C Wang, G Gong, H Su, CW Yu. Efficacy of integrated photovoltaics-air source heat pump systems for application in Central-south China. Renewable and Sustainable Energy Reviews 49(2015)1190–1197

[86] F.J. Aguilar, S. Aledo, P.V. Quiles. Experimental analysis of an air conditioner powered by photovoltaic energy and supported by the grid. Applied Thermal Engineering 123 (2017) 486–497

[87] M Giampaolo, PM Luigi Colombo, R Stefano, F Damiano. Tiles as solar air heater to support a heat pump for residential air conditioning. Applied Thermal Engineering 102 (2016) 1412–1421

[88] G Wang, Z Quan, Y Zhao, P Xu, C Sun. Experimental study of a novel PV/T- air composite heat pump hot water system. Energy Procedia 70 (2015) 537 – 543

[89] L Ni, D Qv, R Shang, Y Yao, F Niu, W Hu. Experimental study on performance of a solar-air source heat pump system in severe external conditions and switchover of different functions. Applied Thermal Engineering 100 (2016) 434–452

[90] F Niu, L Ni, Y Yao, Y Yu, H Li. Performance and thermal charging/discharging features of a phase change material assisted heat pump system in heating mode. Applied Energy 206 (2017) 784-792

[91] S.K. Chaturvedi, T.M. Abdel-Salam, S.S. Sreedharan, F.B. Gorozabel. Two-stage direct expansion solar-assisted heat pump for high temperature applications. Applied Thermal Engineering 29 (2009) 2093–2099

[92] C Wang, G Gong, H Su, Chuck Wah Yu. Dimensionless and thermodynamic modelling of integrated photovoltaics–air source heat pump systems. Solar Energy 118 (2015) 175–185

[93] N Omid, MCJ Luis, SF Alan. Review of computer models of air-based curtainwall-integrated PV/T collectors. Renewable and Sustainable Energy Reviews 63 (2016) 102-117

[94] J Wu, C Chen, S Pan, J Wei, T Pan, Y Wei, Y Wang, X Wang, and J Su. Experimental Study of the Performance of Air Source Heat Pump Systems Assisted by Low-Temperature Solar-Heated Water Advances in Mechanical Engineering Volume 2013, Article ID 843013, 8 pages. http://dx.doi.org/10.1155/2013/843013

[95] J Ji, J Cai, W Huang, Y Feng. Experimental study on the performance of solar-assisted multi-functional heat pump based on enthalpy difference lab with solar simulator. Renewable Energy 75 (2015) 381-388.

[96] J Cai, J Ji, Y Wang, W Huang. Operation characteristics of a novel dual source multi-functional heat pump system under various working modes. Applied Energy 194 (2017) 236–246.

[97] E Mohamed, S Riffat, S Omer. Low-temperature solar-plate-assisted heat pump: a developed design for domestic applications in cold climate. International journal of Refrigeration 81 (2017) 134-150

[98] WZ Huang, J Ji, N Xu, GQ Li. Frosting characteristics and heating performance of a direct-expansion solar-assisted heat pump for space heating under frosting conditions. Applied Energy 171 (2016) 656-666

[99] WZ Huang, T Zhang, J Ji, N Xu. Numerical study and experimental validation of a direct-expansion solar-assisted heat pump for space heating under frosting conditions. Energy and Buildings 185 (2019) 224-238

[100] JB Long, RC Zhang, J Lu, F Xu. Heat transfer performance of an integrated solar-air source heat pump evaporator. Energy Conversion and Management 184 (2019) 626-635

[101] Y Liu, J Ma, GH Zhou, C Zhang, WL Wan. Performance of a solar air composite heat source heat pump system. Renewable Energy 87 (2016) 1053-1058

[102] S. Dubey, G.S. Sandhu, G.N. Tiwari, Analytical expression for electrical efficiency of PV/T hybrid air collector, Appl. Energy 86 (2009) 697–705.

[103] ML Qu, JB Chen, LJ Nie, FS Li, Q Yu, T Wang. Experimental study on the operating characteristics of a novel photovoltaic/thermal integrated dual-source heat source heat pump water heating system. Applied Thermal Engineering 94 (2016) 819-826

[104] G Besagni, L Croci, R Nesa, L Molinaroli. Field study of a novel solar-assisted dual-source multifunctional heat pump. Renewable Energy 132 (2019) 1185-1215

[105] SX Lu, RB Liang, JL Zhang, C Zhou. Performance improvement of solar photovoltaic/thermal heat pump system in winter by employing vapor injection cycle. Applied Thermal Enginnering 155 (2019) 135-146

[106] G Clara. Environmental impact assessments of hybrid photovoltaic–thermal (PV/T) systems – A review. Renewable and Sustainable Energy Reviews55(2016)234–239

[107] G Clara, A Inger, GH Anne. Solar energy for net zero energy buildings – A comparison between solar thermal, PV and photovoltaic–thermal (PV/T) systems Solar Energy 122 (2015) 986–996

[108] K. Kramer, H. Helmers, The interaction of standards and innovation: Hybrid photovoltaic–thermal collectors, Solar Energy, 98, Part C (2013) 434-439

[109] Solar Keymark Network, Specific CEN Keymark Scheme Rules for Solar Thermal Product version 21.00, J.E. Nielsen (Ed.), CEN Certification, 2013